\documentclass[11pt]{article}

\usepackage{mathrsfs,amsbsy,amssymb,latexsym,amsfonts,amsmath}
\usepackage[dvipdfm]{graphicx,color}
\usepackage{wrapfig}
\usepackage{amsbsy}
\usepackage{alltt}

\allowdisplaybreaks[1]

\makeatletter
	
	\@addtoreset{equation}{section}
\makeatother

\makeatletter

\@addtoreset{figure}{section}
\makeatother

\begin{document}

\rightline{KUNS-2521} 
\rightline{OIQP-14-9}


\begin{center}
{\Large \bf A new mechanism of realizing inflationary universe \\ 
\Large \bf with recourse to backreaction of quantized free fields \\ 
-- Inflation without inflaton -- \\}
\end{center}




\centerline{Yoshinobu {\sc Habara}$^{a}$, Hikaru {\sc Kawai}$^{b}$ and Masao {\sc Ninomiya}$^a$}
\medskip
\medskip
\centerline{\small $^a$Okayama Institute for Quantum Physics, 1-9-1 Kyoyama, Okayama 700-0015, Japan}
\centerline{\small $^b$Department of Physics, Kyoto University, Kyoto 606-8502, Japan}
\medskip
\medskip

\noindent
{\small {\bf E-mail:} habara@yukawa.kyoto-u.ac.jp, \\ \hspace{13.25mm} hkawai@gauge.scphys.kyoto-u.ac.jp, \\ \hspace{13.25mm} msninomiya@gmail.com}


\begin{abstract}
It is shown that the inflationary era in early universe is realized due to the effect of backreaction of quantized matter fields. In fact we start by quantizing a free scalar field in the Friedmann-Robertson-Walker space-time, and the field is fluctuating quantum mechanically around the bottom of the mass potential. We then obtain the vacuum expectation value of the energy density of the scalar field as a functional of the scale factor $a(t)$ of the universe. By plugging it into the Einstein equation, a self-consistent equation is established in which the matter fields determine the time evolution of the universe. We solve this equation by setting few conditions and find the following solution: the universe expands \`{a} la de Sitter with e-folding number $\gtrsim 60$ and then it turns to shrink with a decreasing Hubble parameter $H(t)$ which rapidly goes to zero.
\end{abstract}





\section{\label{1}Introduction}

It is believed by most astro-cosmologists as well as astro-particle physicists that the only solution to resolve the flatness and horizon problems in the universe is to think of exponential expansion called inflation~\cite{rf:guth} right after birth of our universe. In recent years several observational results to confirm the inflation were reported~\cite{rf:wmap,rf:planck,rf:bicep2}. In particular, the most important evidence in the present days to provide data is the temperature fluctuation of the cosmic microwave background radiation (CMB) which shows us quantitatively that there was the era of the inflation in the early universe. Using this data we can investigate the mechanism of the inflationary expansion in detail. The CMB is the photon emitted in the recombination era of which the wavelength is very long so that CMB reaches us without much scattering afterwards. Therefore CMB carries to us the information of the early universe. The most peculiar properties of CMB is that it is homogeneous and uniform with very high accuracy and furthermore the spectrum agrees very well with the black body radiation with the temperature 2.725 [K] while the temperature fluctuation $\frac{\delta T}{T}$ is of the order of $10^{-5}$. This temperature fluctuation is believed to come from the gravitational potential $\Phi$ through Sachs-Wolfe effect $\frac{\delta T}{T}=-3\Phi$~\cite{rf:SW-effect}. In the usual inflation models the gravitational potential $\Phi$ is believed to be created when the hypothetical scalar particle(s) called inflaton(s) fluctuates quantum mechanically during rolling down classically along with the potential. There are many models proposed for creation mechanism, such as slow-roll inflation~\cite{rf:slow-roll}, chaotic inflation~\cite{rf:chaotic} etc.

In our previous papers~\cite{rf:ours}, we show that the temperature fluctuation of CMB can be explained as the vacuum fluctuation of a large number of matter fields. Here the matter fields are originated from the string excited states and/or the Kaluza-Klein modes on the compactified space. There appear very many kinds of fields in the exponentially expanding universe, where the Hubble parameter plays the role of temperature. These fields fluctuate around the bottom of the mass potential, and generate the energy density of fluctuation. Finally the fluctuation creates the gravitational potential $\Phi$ through the Einstein equation. Although each fluctuation due to single field is very small, but enormous number of species of fields contribute and it turns out to be summed up to a huge fluctuation compatible to the present observational value. Furthermore it is possible to determine the radii of the compactified spaces, as well as the string coupling constant by comparison with the observational values of CMB. The very merit of our theory is in the fact that we do not need to put any assumption or fine-tuning on the form of the matter fields potential, but we simply consider free fields. However in our above described theory there still are several problems to be resolved. In the first place the spectral index $n_s$ which indicates the distance dependence of the fluctuation or, in other words, the 2-point correlation function of the fields exceeds one, i.e. $n_s\geq 1$ in our model, although the observed value of the Planck is $n_s\simeq 0.96$~\cite{rf:planck}. As was pointed out in our foregoing papers \cite{rf:ours} the reason why we obtained $n_s\geq 1$ is in the fact that we assumed the background metric is exactly the de Sitter space-time. In order to explain the experimental value of the spectral index, we may have to introduce a time dependent Hubble parameter $H(t)$, something like $\frac{\dot{H}}{H^2}\sim -0.01$. 

So far we have not discussed the origin of the inflation, and simply assumed the de Sitter space-time. In this paper we point out the possibility that the inflation is caused by the vacuum energy of the many fields in the expanding universe.

It is the purpose of the present article that we present a self-consistent theory of the time evolution of the universe by incorporating the backreaction of the quantized matter fields\footnote{The effect of backreaction has been taken into account, in \cite{rf:kaya}. However in our analysis, the method of approximation and the choice of parameters are different from them so that the results and interpretation are really different.}. The main idea of the present paper follows our pervious articles \cite{rf:ours}, except that we do not assume de Sitter background and incorporate the backreaction of the matter fields. In the early universe we consider $N$ species of the free scalar fields. The time evolution of the early universe is assumed to follow the classical Einstein equation, 
\begin{align}
	R_{\mu \nu}-\frac{1}{2}g_{\mu \nu}R
	=-8\pi G\sum_{i=1}^N\langle T_{\mu \nu}^{(i)}\rangle .
	\label{1-einstein.eq}
\end{align}
On the right hand side the energy-momentum tensor of the species $i$ of the matter fields is regarded as the vacuum expectation value $\langle T_{\mu \nu}^{(i)}\rangle$, in the metric $g_{\mu \nu}$. When one expresses $\langle T_{\mu \nu}^{(i)}\rangle$ by a functional of arbitrary metric $g_{\mu \nu}$, $\langle T_{\mu \nu}^{(i)}\rangle [g]$, the Einstein equation (\ref{1-einstein.eq}) describes the time evolution of $g_{\mu \nu}$ caused by the backreaction from the matter fields. Here we take the vacuum of the Bunch-Davies type in the calculation of $\langle T_{\mu \nu}^{(i)}\rangle [g]$. As we will see $\langle T_{\mu \nu}^{(i)}\rangle [g]$is proportional to $H^4(t)$. Therefore in the early universe, where $H(t)$ is large, we can ignore the initial state dependence of the energy-momentum tensor and simply take $\langle T_{\mu \nu}^{(i)}\rangle [g]$.

We mainly use the Eikonal approximation to make solutions of the equation of motion of the matter fields. Then we perform the 1-loop momentum integration to calculate the correlation functions. By so doing the eq. (\ref{1-einstein.eq}) becomes a second order differential equation for $H(t)$.

The present paper is organized as follows. In section \ref{2}, we review briefly the Eikonal approximation. Then we investigate the solution of matter field of mass $m$ in two cases, that is, for the cases of $m\ll H$ and $m\gg H$, and evaluate $\langle T_{\mu \nu}\rangle$ for each cases. In section \ref{3} we assume that there exist both light and heavy particles in the universe and search for de Sitter-like expanding solutions of eq. (\ref{1-einstein.eq}), where $H(t)$ is almost constant. Then we examine the solutions when the Hubble parameter is very small compared to that of the de Sitter-like expanding era. From this analysis we will see that the Hubble parameter becomes zero and the de Sitter era ends for a certain region of the initial conditions. Section \ref{4} is devoted to conclusion and outlook. In Appendix \ref{app-airy} we describe in detail the calculation by using the Airy function around boundary between UV and IR regimes in the Eikonal approximation. In Appendix \ref{app-damping} we show the calculation of the correlation functions using the solution in the IR regime in the Eikonal approximation.

\section{\label{2}Scalar field in the Friedmann-Robertson-Walker space-time and the Eikonal approximation}

It is legitimate to assume that the background metric is homogeneous and uniform as are indicated from CMB data. We thus start with investigating the one species of the scalar field described by the action 
\begin{align}
	S=\int d^nx\> \frac{a^{n-1}}{2}\left\{\dot{\phi}^2
	-\frac{1}{a^2}\left(\nabla \phi \right)^2-m^2\phi^2\right\}
	\label{2-action}
\end{align}
in the Friedmann-Robertson-Walker space-time 
\begin{align}
	g_{\mu \nu}(x)=\left( \begin{array}{cccc}
	1 & 0 & 0 & 0 \\
	0 & -a^2(t) & 0 & 0 \\
	0 & 0 & -a^2(t) & 0 \\
	0 & 0 & 0 & -a^2(t) 
	\end{array} \right). 
	\label{2-FRW.metric}
\end{align}
For the later convenience to apply the dimensional regularization, the space-time dimension in eq. (\ref{2-action}) is taken to be $n=4-\epsilon$ with $0<\epsilon \ll 1$. In terms of the co-moving time 
\begin{align}
	d\tau =\frac{dt}{a}, 
	\label{2-comoving.time}
\end{align}
we write the differentiation as 
\begin{align}
	F^{\prime}=\frac{d}{d\tau}F=a\frac{d}{dt}F=a\dot{F}, 
	\label{2-comoving.time-diff}
\end{align}
and then the Fourier mode $\chi_k(\tau )$ of the rescaled scalar field 
\begin{align}
	\chi (\tau ,\vec{x})\equiv a^{\frac{n-2}{2}}(t)\phi (t,\vec{x})
	\label{2-comoving.field}
\end{align}
obeys the following equation of motion: 
\begin{align}
	\chi_k^{\prime \prime}+\left(k^2-V[a]\right)\chi_k=0.
	\label{2-eom}
\end{align}
Here the potential $V[a]$ is defined by 
\begin{align}
	V[a] 
	& \equiv \left(1-\frac{\epsilon}{2}\right)\frac{a^{\prime \prime}}{a}
	-\epsilon \left(\frac{1}{2}-\frac{\epsilon}{4}\right)
	\left(\frac{a^{\prime}}{a}\right)^2-m^2a^2 \nonumber \\
	& =\left(1-\frac{\epsilon}{2}\right)a^2\left(2H^2+\dot{H}\right)
	-\epsilon \left(\frac{1}{2}-\frac{\epsilon}{4}\right)a^2H^2-m^2a^2, 
	\label{2-potential}
\end{align}
where $H(t)$ is the Hubble parameter 
\begin{align}
	H(t)=\frac{\dot{a}(t)}{a(t)}.
	\label{2-hubble.parameter}
\end{align}
The quantity we are interested in is the $(0,0)$-component of the energy-momentum tensor as a functional of $a(t)$ (or $H(t)$), 
\begin{align}
	\langle T_{00}\rangle [a] 
	& =\frac{\mu^{4-n}}{2}\left\langle\dot{\phi}^2
	+\frac{1}{a^2}\left(\nabla \phi \right)^2+m^2\phi^2\right\rangle 
	\nonumber \\	
	& =\frac{\mu^{\epsilon}}{2a^{4-\epsilon}}\bigg\{
	\left\langle \left(\chi^{\prime}\right)^2\right\rangle 
	-\left(1-\frac{\epsilon}{2}\right)aH
	\left\langle \chi^{\prime}\chi +\chi \chi^{\prime}\right\rangle 
	+\left(1-\frac{\epsilon}{2}\right)^2a^2H^2
	\left\langle \chi^2\right\rangle \nonumber \\
	& \qquad \qquad \quad 
	+\left\langle \left(\nabla \chi \right)^2\right\rangle 
	+m^2a^2\left\langle \chi^2\right\rangle \bigg\}.
	\label{2-energy-momentum.tensor}
\end{align}
Here a mass scale $\mu$ is introduced to keep the mass dimension of the energy-momentum tensor.

We solve the equation of motion (\ref{2-eom}) with recourse to the Eikonal approximation by putting 
\begin{align}
	\chi_k=Ae^{-if}.
	\label{2-eikonal}
\end{align}
Then by using the boundary condition 
\begin{align}
	\chi_k\sim \frac{1}{\sqrt{k}}e^{-ik\tau},\quad 
	(\tau \to -\infty )
	\label{2-initial.condition}
\end{align}
and the assumption of the Eikonal approximation 
\begin{align}
	\left(f^{\prime}\right)^2\gg \frac{A^{\prime \prime}}{A}, 
	\label{2-eikonal.condition}
\end{align}
(\ref{2-eom}) has the solution 
\begin{align}
	& \chi_k=\frac{1}{\sqrt{2f^{\prime}}}e^{-if}, \quad
	f^{\prime}=\sqrt{k^2-V[a]}.
	\label{2-eikonal.solution}
\end{align}
Here the Wronskian is used to normalize the solution: 
\begin{align}
	\chi_k(\chi_k^{\ast})^{\prime}-\chi_k^{\ast}\chi_k^{\prime}=i.
	\label{2-wronskian}
\end{align}
The general solution for $\chi (\tau ,\vec{x})$ is given by 
\begin{align}
	\chi (\tau ,\vec{x})
	=\int \frac{d^{3-\epsilon}\vec{k}}{(2\pi )^{3-\epsilon}}
	\left\{\frac{1}{\sqrt{2f^{\prime}(k,\tau )}}
	e^{-if(k,\tau )-i\vec{k}\cdot \vec{x}}a_{\vec{k}}+\text{c.c.}\right\}
	\label{2-general.solution}
\end{align}
in the oscillatory region, 
\begin{align}
	k^2-V>0.
	\label{2-oscillatory.region}
\end{align}
We then quantize the field $\chi$ canonically by using the commutation relation 
\begin{align}
	[\chi (\tau ,\vec{x}),\chi^{\prime}(\tau ,\vec{y})]
	=i\delta^{3-\epsilon}(\vec{x}-\vec{y}).
	\label{2-commutation.relation}
\end{align}
The vacuum state is defined as usual 
\begin{align}
	a_{\vec{k}}|0\rangle =0
	\label{2-vacuum}
\end{align}
so that the vacuum expectation value of an operator $\mathcal{O}$ is given by $\left\langle \mathcal{O} \right\rangle=\langle 0|\mathcal{O} |0\rangle$. 

In the following we consider the situation where $\dot{H}(t)$ and $\ddot{H}(t)$ are much smaller than $H(t)$ itself, 
\begin{align}
	\frac{\dot{H}}{H^2}\ll 1 \quad \text{and}
	\quad \frac{\ddot{H}}{H^3}\ll 1, 
	\label{2-slow.develop}
\end{align}
which means that $H(t)$ and $\dot{H}(t)$ vary much slower than the time evolution of the universe.

\subsection{\label{2.1}Energy density for the case of a large Hubble parameter: $H(t)\gg m$}

In this subsection we evaluate the energy density $\rho$ of a scalar field when Hubble parameter is much larger than mass; $H(t)\gg m$. It is well known that if $H(t)$ is constant $\rho$ is given by 
\begin{align}
	\rho =\left\langle T_{00}\right\rangle =\frac{3}{32\pi^2}H^4
	\left\{1+O\left(\frac{m^2}{H^2}\right)\right\}.
	\label{2-pure.de.Sitter}
\end{align}
Note that this result is obtained without using the Eikonal approximation. 

When $\frac{\dot{H}}{H^2}\ll 1$ and $\frac{\ddot{H}}{H^3}\ll 1$, the potential (\ref{2-potential}) is always positive, i.e. 
\begin{align}
	V[a]>0.
	\label{2-potential.is.positive}
\end{align}
Then the eq. (\ref{2-oscillatory.region}) is not always satisfied, and we should analyze the following three momentum regions to continue the solution $\chi_k\sim e^{-ik\tau}$ in the ultra-violet limit $\tau \to -\infty$ (equivalently $k\to \infty$) into the infrared region: 
\begin{enumerate}
\item[(i)] $\sqrt{V}<k<\infty$ 

In this region the solution is oscillatory as 
\begin{align}
	\chi_k(\tau )=\frac{1}{\sqrt{2}}\frac{1}{\sqrt[4]{k^2-V(\tau )}}
	e^{-i\int_{\tau_k}^{\tau} d\tau^{\prime} \sqrt{k^2-V(\tau^{\prime})}}.
	\label{2-oscillation.solution}
\end{align}
Here $\tau_k$ is determined from $k^2=V(\tau_k)$.
\item[(ii)] $k_0<k<\sqrt{V}$ 

The solution is expressed in terms of the Airy function 
\begin{align}
	\chi_k(\tau )=\frac{e^{-i\frac{\pi}{4}}}{\sqrt{2}}B^{-\frac{1}{6}}
	\left\{iAi\left(B^{\frac{1}{3}}(\tau -\tau_k)\right)
	+Bi\left(B^{\frac{1}{3}}(\tau -\tau_k)\right)\right\}, \quad 
	B\equiv V^{\prime}(\tau_k).
	\label{2-airy.solution}
\end{align}
\item[(iii)] $0\leq k<k_0$ 

The damping and enhancing solutions given by 
\begin{align}
	\chi_k(\tau )=\frac{e^{-i\frac{\pi}{4}}}{\sqrt{2}}
	\frac{1}{\sqrt[4]{V(\tau )-k^2}}\left\{\frac{i}{2}
	e^{-\int_{\tau_k}^{\tau}d\tau^{\prime}\sqrt{V(\tau^{\prime})-k^2}}
	+e^{+\int_{\tau_k}^{\tau}d\tau^{\prime}\sqrt{V(\tau^{\prime})-k^2}}
	\right\}.
	\label{2-damping.solution}
\end{align}
\end{enumerate}
In the above (ii) and (iii), the boundary $k_0$ is chosen to satisfy the validity condition of the Eikonal approximation (\ref{2-eikonal.condition}). In fact, 
\begin{align}
	\left(f^{\prime}\right)^2\gg \frac{A^{\prime \prime}}{A} 
	\quad \Rightarrow \quad 
	\left(\frac{1}{f^{\prime}}\right)^{\prime}\ll 1 
	\quad \Rightarrow \quad 
	k\ll \sqrt{V-\left(\frac{V^{\prime}}{2}\right)^{\frac{2}{3}}}
	\equiv k_0.
	\label{2-airy.boundary}
\end{align}
Here $V^{\prime}$ is obtained from (\ref{2-comoving.time-diff}) and (\ref{2-potential}), 
\begin{align}
	V^{\prime}=\left(1-\frac{\epsilon}{2}\right)
	a^3\left(4H^3+6H\dot{H}+\ddot{H}\right)
	-\epsilon \left(1-\frac{\epsilon}{2}\right)
	a^3\left(H^3+H\dot{H}\right)-2m^2a^3H.
	\label{2-potential-prime}
\end{align}

We next evaluate the vacuum expectation value of the energy density in each region. In the region (i) it is given by 
\begin{align}
	\left\langle T_{00}\right\rangle_{H\gg m}
	& =\frac{1}{32\pi^2}\left(-m^4+2m^2H^2-6H^2\dot{H}-2H\ddot{H}+\dot{H}^2
	\right)\frac{1}{\epsilon} \nonumber \\
	& \quad -\frac{1}{384\pi^2}\frac{1}{2H^2+\dot{H}-m^2} \nonumber \\
	& \qquad \times 
	\Big[
	6\left\{\textstyle \frac{4}{3}H^4
	-2\left(\log 4\pi -\gamma +\frac{1}{3}\right)m^2H^2
	+\left(\log 4\pi -\gamma +\frac{3}{2}\right)m^4
	\right\}\left(2H^2-m^2\right) \nonumber \\
	& \qquad \qquad 
	+6\big\{\textstyle 12\left(\log 4\pi -\gamma -2\right)H^4
	-8\left(\log 4\pi -\gamma -\frac{5}{4}\right)m^2H^2 \nonumber \\
	& \qquad \qquad \qquad \textstyle 
	+\left(\log 4\pi -\gamma -\frac{5}{2}\right)m^4\big\}\dot{H} 
	\nonumber \\
	& \qquad \qquad +12\left\{\textstyle 
	+\left(\log 4\pi -\gamma -1\right)\dot{H}
	+\left(\log 4\pi -\gamma -\frac{5}{3}\right)\left(2H^2-m^2\right)
	\right\}H\ddot{H} \nonumber \\
	& \qquad \qquad 
	+6\left\{\textstyle 4\left(\log 4\pi -\gamma -\frac{7}{4}\right)H^2
	+\left(\log 4\pi -\gamma -\frac{5}{2}\right)m^2\right\}\dot{H}^2 
	\nonumber \\
	& \qquad \qquad \textstyle 
	-6\left(\log 4\pi -\gamma -\frac{3}{2}\right)\dot{H}^3
	+\ddot{H}^2\Big] \nonumber \\
	& \quad 
	-\frac{1}{64\pi^2}\left(-m^4+2m^2H^2-6H^2\dot{H}-2H\ddot{H}+\dot{H}^2
	\right)\log \frac{2H^2+\dot{H}-m^2}{\mu^2}+O(\epsilon ).
	\label{2-em.tensor-oscillation}
\end{align}
Only in this region there appears an ultra-violet divergence which should be renormalized. Indeed in (\ref{2-em.tensor-oscillation}), the $\frac{1}{\epsilon}$ term gives the divergence in dimensional regularization. There are five terms in the coefficient of $\frac{1}{\epsilon}$. The $m^4$ and $m^2H^2$ terms are renormalized into the cosmological constant term $\int d^4x\sqrt{-g}$ and the Einstein-Hilbert action $\int d^4x\sqrt{-g}R$, respectively. The rest three terms are $H^2\dot{H}$, $H\ddot{H}$, and $\dot{H}^2$ and from the dimensional analysis their counter terms are given by $\int d^4x\sqrt{-g}R^2$ and $\int d^4x\sqrt{-g}R_{\mu \nu}R^{\mu \nu}$. In the case of FRW metric (\ref{2-FRW.metric}), these two actions give the following additional terms to the Euler-Lagrange equation: 
\begin{align}
	\int d^4x\sqrt{-g}\>R^2\quad \Rightarrow \quad 
	& 18\left(-6H^2\dot{H}-2H\ddot{H}+\dot{H}^2\right), \nonumber \\
	\int d^4x\sqrt{-g}R_{\mu \nu}R^{\mu \nu}\quad \Rightarrow \quad 
	& 6\left(-6H^2\dot{H}-2H\ddot{H}+\dot{H}^2\right).
	\label{2-counter.term}
\end{align}
Therefore the three coefficients of the $\frac{1}{\epsilon}$ term can be renormalized by these two counter terms. In principle we may introduce finite renormalization for the three terms $m^4$, $m^2H^2$ and $\Big(-6H^2\dot{H}-2H\ddot{H}+\dot{H}^2\Big)$. However, since we have to make the same renormalization to the two cases $H(t)\gg m$ and $H(t)\ll m$ of the single theory, we relegate to argue this problem after the following subsection \ref{2.2} for the case of a small Hubble parameter $H(t)\ll m$.

Then we make use of the minimal subtraction, and expand (\ref{2-em.tensor-oscillation}) for large $H(t)$ assuming $H(t)\gg m,\> \frac{\dot{H}}{H^2}\ll 1$ and $\frac{\ddot{H}}{H^3}\ll 1$: 
\begin{align}
	\left\langle T_{00}\right\rangle_{H\gg m}
	& \simeq \frac{1}{64\pi^2}\Big\{\textstyle -\frac{4}{3}H^4
	-6\left(\log 4\pi -\gamma -\frac{19}{9}\right)H^2\dot{H}
	-2\left(\log 4\pi -\gamma -\frac{5}{3}\right)H\ddot{H} \nonumber \\
	& \qquad \qquad 
	+\left(\log 4\pi -\gamma -\frac{17}{6}\right)\dot{H}^2\Big\} 
	\nonumber \\
	& \quad 
	-\frac{1}{64\pi^2}\left(-6H^2\dot{H}-2H\ddot{H}+\dot{H}^2\right)
	\log \frac{2H^2}{\mu^2}+\cdots .
	\label{2-em.tensor-oscillation-approx}
\end{align}

For the regions (ii) and (iii) the contributions to $\left\langle T_{00}\right\rangle_{H\gg m}$ are described in the appendices \ref{app-airy} and \ref{app-damping} respectively. We here show the combined result of (ii) and (iii) calculated numerically under the conditions $H(t)\gg m$, $\frac{\dot{H}}{H^2}\ll 1$ and $\frac{\ddot{H}}{H^3}\ll 1$: 
\begin{align}
	\left\langle T_{00}\right\rangle_{H\gg m}
	& \simeq 0.01365H^4+0.01333H^2\dot{H}
	+0.0005572H\ddot{H}+0.02139\dot{H}^2+\cdots .
	\label{2-em.tensor-airy+damping}
\end{align}
Note that the coefficient of $H\ddot{H}$ does not remain stationary when we vary the artificial parameter $k_0$, which was introduced to separate two regions (ii) and (iii). However this coefficient is very small compared to the other terms and does not affect the physical result.

\subsection{\label{2.2}Energy density for the case of a small Hubble parameter: $H(t)\ll m$}

In this subsection we derive the energy density in the case of a small Hubble parameter, $H(t)\ll m$. In the same manner as the previous subsection \ref{2.1} we calculate the vacuum expectation value of the energy density. Again we assume that the change of the Hubble parameter is fairly slow, 
\begin{align*}
	\frac{\dot{H}}{H^2}\ll 1,\quad \frac{\ddot{H}}{H^3}\ll 1.
\end{align*}
In this case the potential term (\ref{2-potential}) of the equation of motion (\ref{2-eom}) is always negative, 
\begin{align}
	V[a]<0.
	\label{2-potential.is.negative}
\end{align}
Then the solution of the Eikonal approximation (\ref{2-eikonal.solution}) is expressed as 
\begin{align*}
	f^{\prime}=\sqrt{k^2+\left(-V[a]\right)}.
\end{align*}
Apparently any turning point does not exist so that the solution (\ref{2-eikonal.solution}) is valid in the whole region of $0\leq k<\infty$.

Thus we obtain the energy density as 
\begin{align}
	\left\langle T_{00}\right\rangle_{H\ll m}
	& =\frac{1}{32\pi^2}\left(-m^4+2m^2H^2-6H^2\dot{H}-2H\ddot{H}+\dot{H}^2
	\right)\frac{1}{\epsilon} \nonumber \\
	& \quad +\frac{1}{384\pi^2}\frac{1}{m^2-2H^2-\dot{H}} \nonumber \\
	& \qquad \times 
	\Big[
	-6\left\{\textstyle \left(\log 4\pi -\gamma +\frac{3}{2}\right)m^4
	-2\left(\log 4\pi -\gamma +\frac{1}{3}\right)m^2H^2+\frac{4}{3}H^4
	\right\}\left(m^2-2H^2\right) \nonumber \\
	& \qquad \qquad 
	+6\big\{\textstyle \left(\log 4\pi -\gamma -\frac{5}{2}\right)m^4
	-8\left(\log 4\pi -\gamma -\frac{5}{4}\right)m^2H^2 \nonumber \\
	& \textstyle \qquad \qquad \qquad 
	+12\left(\log 4\pi -\gamma -2\right)H^4\big\}\dot{H} \nonumber \\
	& \qquad \qquad 
	+12\left\{\textstyle 
	-\left(\log 4\pi -\gamma -\frac{5}{3}\right)\left(m^2-2H^2\right)
	+\left(\log 4\pi -\gamma -1\right)\dot{H}\right\}H\ddot{H} \nonumber \\
	& \qquad \qquad 
	+6\left\{\textstyle \left(\log 4\pi -\gamma -\frac{5}{2}\right)m^2
	+4\left(\log 4\pi -\gamma -\frac{7}{4}\right)H^2\right\}\dot{H}^2 
	\nonumber \\
	& \qquad \qquad \textstyle 
	-6\left(\log 4\pi -\gamma -\frac{3}{2}\right)\dot{H}^3
	+\ddot{H}^2\Big] \nonumber \\
	& \quad 
	-\frac{1}{64\pi^2}\left(-m^4+2m^2H^2-6H^2\dot{H}-2H\ddot{H}+\dot{H}^2
	\right)\log \frac{m^2-2H^2-\dot{H}}{\mu^2}+O(\epsilon ).
	\label{2-em.tensor-damping}
\end{align}
We then make the renormalization by using the minimal subtraction, and the expansion assuming $H(t)\ll m,\> \frac{\dot{H}}{H^2}\ll 1$ and $\frac{\ddot{H}}{H^3}\ll 1$ gives the result, 
\begin{align}
	\left\langle T_{00}\right\rangle_{H\ll m}
	& \simeq \frac{1}{64\pi^2}\Big\{\textstyle 
	-\left(\log 4\pi -\gamma +\frac{3}{2}\right)m^4
	+2\left(\log 4\pi -\gamma -\frac{2}{3}\right)m^2H^2 \nonumber \\
	& \qquad \qquad \textstyle +\frac{2}{3}H^4
	-6\left(\log 4\pi -\gamma -\frac{19}{9}\right)H^2\dot{H}
	-2\left(\log 4\pi -\gamma -\frac{5}{3}\right)H\ddot{H} \nonumber \\
	& \qquad \qquad \textstyle 
	+\left(\log 4\pi -\gamma -2\right)\dot{H}^2\Big\} \nonumber \\
	& \quad 
	-\frac{1}{64\pi^2}\left(-m^4+2m^2H^2-6H^2\dot{H}-2H\ddot{H}+\dot{H}^2
	\right)\log \frac{m^2}{\mu^2}+\cdots .
	\label{2-em.tensor-damping-approx}
\end{align}

Hereby we next make use of the finite renormalization for the cosmological constant term $\int d^4x\sqrt{-g}$ and the Einstein-Hilbert action $\int d^4x\sqrt{-g}R$ in such a way that $m^4$ and $m^4\log \frac{m^2}{\mu^2}$ terms and $m^2H^2$ and $m^2H^2\log \frac{m^2}{\mu^2}$ terms are set to zero. The reason for so doing is because the energy density for small values of $H(t)$ should be zero. Although we should add the same counter terms to (\ref{2-em.tensor-oscillation-approx}), such terms are negligible when $H(t)$ is much larger than $m$. Furthermore we may make a finite renormalization for $\int d^4x\sqrt{-g}R^2$ and $\int d^4x\sqrt{-g}R_{\mu \nu}R^{\mu \nu}$ which generates an additional term $\frac{\lambda}{64\pi^2}\left(-6H^2\dot{H}-2H\ddot{H}+\dot{H}^2\right)$ to (\ref{2-em.tensor-damping-approx}).

\section{\label{3}Solution of the Einstein equation including backreaction}

So far we have obtained the energy density of a single scalar field for two cases. The explicit form for the case of a large Hubble parameter is given by 
\begin{align}
	\left\langle T_{00}\right\rangle_{H\gg m} 
	& =\frac{1}{64\pi^2}\left(-\frac{4}{3}H^4+\frac{38}{3}H^2\dot{H}
	+\frac{10}{3}H\ddot{H}-\frac{17}{6}\dot{H}^2\right) \nonumber \\
	& \quad +\frac{1}{64\pi^2}
	\left(\log 4\pi -\gamma -\log \frac{2H^2}{\mu^2}+\lambda \right)
	\left(-6H^2\dot{H}-2H\ddot{H}+\dot{H}^2\right) \nonumber \\
	& \quad +0.01365H^4+0.01333H^2\dot{H}+0.0005572H\ddot{H}
	+0.02139\dot{H}^2
	\label{3-energy.density-right}
\end{align}
which is the sum of (\ref{2-em.tensor-oscillation-approx}) and (\ref{2-em.tensor-airy+damping}). For the case of a small Hubble parameter we have 
\begin{align}
	\left\langle T_{00}\right\rangle_{H\ll m} 
	& =\frac{1}{64\pi^2}\left(\frac{2}{3}H^4+\frac{38}{3}H^2\dot{H}
	+\frac{10}{3}H\ddot{H}-2\dot{H}^2\right) \nonumber \\
	& \quad +\frac{1}{64\pi^2}
	\left(\log 4\pi -\gamma -\log \frac{m^2}{\mu^2}+\lambda \right)
	\left(-6H^2\dot{H}-2H\ddot{H}+\dot{H}^2\right).
	\label{3-energy.density-heavy}
\end{align}

We next consider the Einstein equation assuming that there are many fields with various masses. Then the total energy density is written as 
\begin{align}
	\left\langle T_{00}\right\rangle 
	=N_1\left\langle T_{00}\right\rangle_{H\gg m}
	+N_2\left\langle T_{00}\right\rangle_{H\ll m},
	\label{3-total.energy0}
\end{align}
where $N_1$ and $N_2$ stand for the numbers of the fields whose masses are lighter and heavier than the Hubble parameter, respectively. The ratio of $N_1$ to $N_2$ may vary as the Hubble parameter evolves in time. However, the coefficients of each terms in (\ref{3-energy.density-right}) and (\ref{3-energy.density-heavy}) are roughly the same, so that we may replace $\left\langle T_{00}\right\rangle_{H\gg m}$ and $\left\langle T_{00}\right\rangle_{H\ll m}$ with their average. Then (\ref{3-total.energy0}) becomes 
\begin{align*}
	\left\langle T_{00}\right\rangle 
	=\frac{N}{2}\Big(\left\langle T_{00}\right\rangle_{H\gg m}
	+\left\langle T_{00}\right\rangle_{H\ll m}\Big), 
\end{align*}
where $N$ is the total number of the species. Then the $(0,0)$-component of the Einstein equation reads 
\begin{align}
	R_{00}-\frac{1}{2}g_{00}R=-8\pi GN\times \frac{1}{2}
	\Big(\left\langle T_{00}\right\rangle_{H\gg m}
	+\left\langle T_{00}\right\rangle_{H\ll m}\Big).
	\label{3-einstein.eq.0}
\end{align}
Rewriting this equation after plugging (\ref{3-energy.density-right}) and (\ref{3-energy.density-heavy}), we obtain the equation of describing the time evolution of the Hubble parameter $H(t)$: 
\begin{align}
	-3H^2 & =-8\pi GN\times \frac{1}{2}\Bigg\{
	\frac{1}{64\pi^2}\left(-\frac{2}{3}H^4+\frac{76}{3}H^2\dot{H}
	+\frac{20}{3}H\ddot{H}-\frac{29}{6}\dot{H}^2\right) \nonumber \\
	& \qquad \qquad \qquad \quad +\frac{1}{32\pi^2}
	\left(\log 4\pi -\gamma -\log \frac{\sqrt{2}Hm}{\mu^2}+\lambda \right)
	\left(-6H^2\dot{H}-2H\ddot{H}+\dot{H}^2\right) \nonumber \\
	& \qquad \qquad \qquad \quad 
	+0.01365H^4+0.01333H^2\dot{H}+0.0005572H\ddot{H}+0.02139\dot{H}^2
	\Bigg\}, 
	\label{3-einstein.eq}
\end{align}
where $m$ is the typical mass of the fields heavier than the Hubble parameter. A comment is in order: The mass scale $\mu$ is originally arbitrary, but as a cut-off scale in the general relativity, field theories and/or string theory, it may be natural to choose Planck scale $m_{pl}$ or string scale $m_s$.

\subsection{\label{3.1}Inflationary solution}

In this subsection we analytically solve the above eq. (\ref{3-einstein.eq}) by using linear approximation around the de Sitter solution. It is clear that the eq. (\ref{3-einstein.eq}) has the solution, where $H(t)$ is a constant, 
\begin{align}
	H(t)=\frac{1}{\sqrt{\frac{8\pi GN}{3}\times \frac{1}{2}
	\left\{-\frac{1}{64\pi^2}\frac{2}{3}+0.01365\right\}}}
	\simeq 4.353\times \frac{m_{pl}}{\sqrt{N}}\equiv H_0, 
	\label{3-pure.de.sitter}
\end{align}
which is nothing but the de Sitter solution. We next consider a fluctuation $\tilde{H}$ around this solution $H_0$, 
\begin{align}
	H=H_0+\tilde{H}.
	\label{3-pure.de.sitter-fluctuation}
\end{align}
In the linear approximation the eq. (\ref{3-einstein.eq}) becomes 
\begin{align}
	r\ddot{\tilde{H}}+\frac{q}{\sqrt{p}}\dot{\tilde{H}}+2\tilde{H}=0.
	\label{3-einstein.eq.2}
\end{align}
where $p,\> q$ and $r$ are constants given by 
\begin{align}
	p & =\frac{8\pi N}{3m_{pl}^2}\times \frac{1}{2}
	\left\{-\frac{1}{64\pi^2}\frac{2}{3}+0.01365\right\}, \nonumber \\
	q & =\frac{8\pi N}{3m_{pl}^2}\times \frac{1}{2}
	\left\{\frac{1}{64\pi^2}\frac{76}{3}
	-\frac{6}{32\pi^2}\left(\log 4\pi -\gamma 
	-\log \frac{\sqrt{2}H_0m}{\mu^2}+\lambda \right)+0.01333\right\}, 
	\nonumber \\
	r & =\frac{8\pi N}{3m_{pl}^2}\times \frac{1}{2}
	\left\{\frac{1}{64\pi^2}\frac{20}{3}
	-\frac{2}{32\pi^2}\left(\log 4\pi -\gamma 
	-\log \frac{\sqrt{2}H_0m}{\mu^2}+\lambda \right)+0.0005572\right\}.
	\label{3-einstein.eq.2-constants}
\end{align}
The general solution reads 
\begin{align}
	& H=H_0\left(1-C_1e^{h_1t}-C_2e^{h_2t}\right),\qquad C_1,C_2\ll 1, 
	\nonumber \\
	& h_1,h_2
	=\frac{-\frac{q}{\sqrt{p}}\pm \sqrt{\frac{q^2}{p}-8r}}{2r}.
	\label{3-einstein.eq-begin-solution}
\end{align}

\begin{center}
\begin{figure}[htbp]
	\includegraphics[width=8cm]{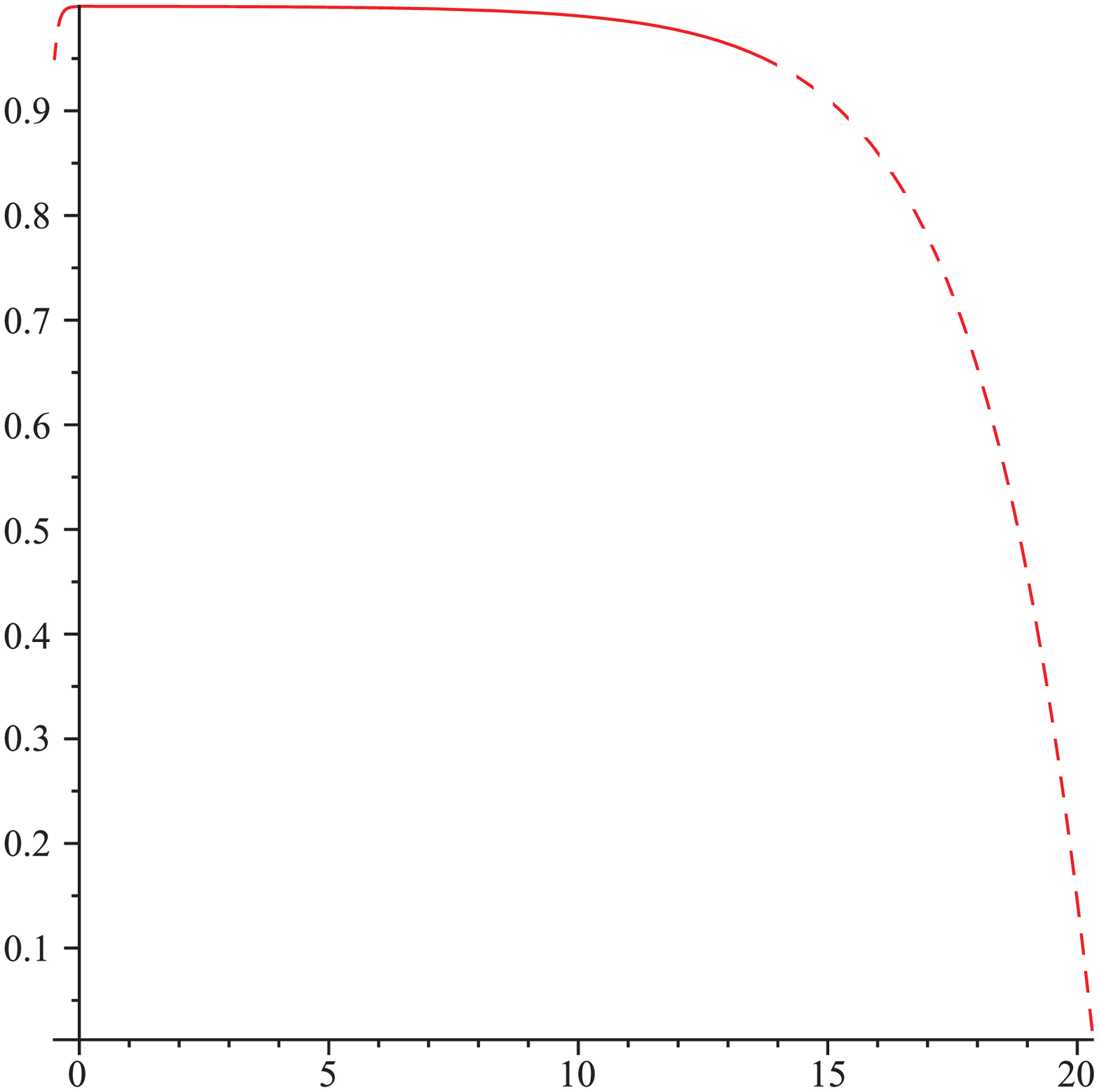}
	\includegraphics[width=8cm]{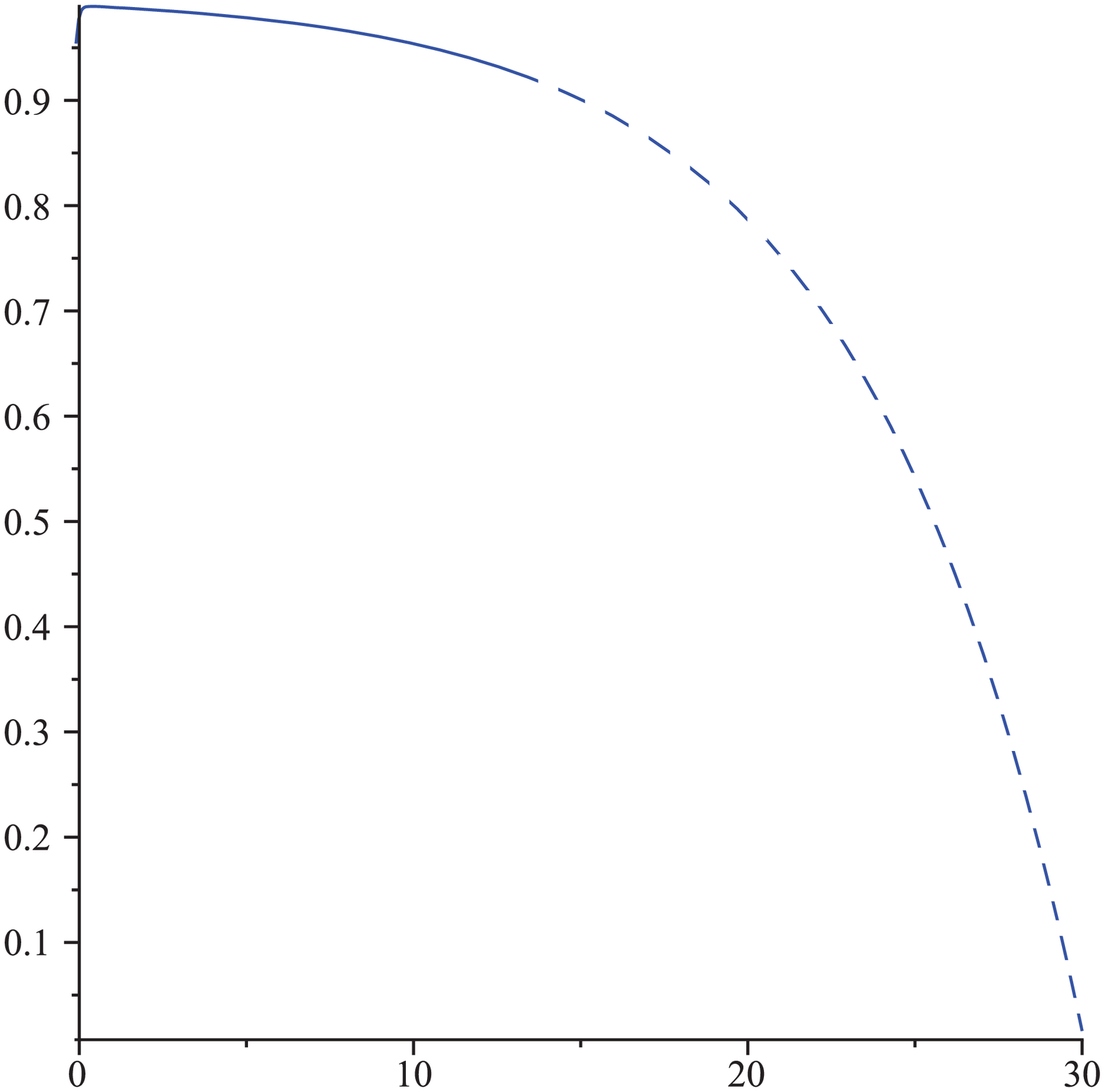}
	\put(-350,0){$\frac{m_{pl}}{\sqrt{N}}t$}
	\put(-120,0){$\frac{m_{pl}}{\sqrt{N}}t$}
	\put(-480,120){$\frac{H(t)}{H_0}$}
	\put(-380,100){(a)}
	\put(-360,100){$\lambda =0$}
	\put(-360,80){$C_1=10^{-4}$}
	\put(-360,60){$C_2=10^{-4}$}
	\put(-150,100){(b)}
	\put(-130,100){$\lambda =25$}
	\put(-130,80){$C_1=10^{-2}$}
	\put(-130,60){$C_2=10^{-2}$}
	\label{fig:energy-backreaction_inflation1}
	\caption{The time evolution of the Hubble parameter $H(t)$: 
	The inflation.}
\end{figure}
\end{center}

Let us turn to investigate the behavior of the solution (\ref{3-einstein.eq-begin-solution}) by using observational values. If we assume that the matter field potential is about the forth power of the GUT scale, $V\sim 10^{65}\text{ [GeV$^4$]}$~\cite{rf:bicep2}, we obtain 
\begin{align}
	H_0^2=\frac{8\pi G}{3}V\quad \Rightarrow \quad 
	H_0\simeq 10^{14}\text{ [GeV]}.
	\label{3-hubble}
\end{align}
Consequently, from (\ref{3-pure.de.sitter}), the number of species $N$ is evaluated as 
\begin{align}
	N\simeq 10^{11}.
	\label{3-number.of.species}
\end{align}
Let us, for example, choose $\mu$ and $m$ in (\ref{3-einstein.eq}) as 
\begin{align}
	& \mu =m_s\simeq 10^{18}\text{ [GeV]}, \nonumber \\
	& m=10^{16}\text{ [GeV]}, 
	\label{3-scales}
\end{align}
where $m_s$ is the string scale. In the case of $\lambda =0$, in which we do not make an additional finite renormalization to the minimal subtraction, the solution (\ref{3-einstein.eq-begin-solution}) becomes 
\begin{align}
	H(t) & \simeq H_0\left(1-C_1e^{-12.48\frac{m_{pl}}{\sqrt{N}}t}
	-C_2e^{0.4527\frac{m_{pl}}{\sqrt{N}}t}\right) \nonumber \\
	& \simeq H_0\left(1-C_1e^{-2.866H_0t}-C_2e^{0.1040H_0t}\right).
	\label{3-einstein.eq-begin-solution1}
\end{align}
If $C_2$ satisfies 
\begin{align*}
	C_2\lesssim 10^{-4}, 
\end{align*}
the duration of the inflation becomes 
\begin{align*}
	\Delta t\gtrsim \frac{60}{H_0}, 
\end{align*}
i.e. the e-folding number exceeds $60$. In Fig.~\ref{fig:energy-backreaction_inflation1}(a) we depict the time evolution of the Hubble parameter with the choices $C_1=10^{-4}$ and $C_2=10^{-4}$. The present approximation $\frac{\dot{H}}{H^2}\ll 1$ and $\frac{\ddot{H}}{H^3}\ll 1$ is valid in the region 
\begin{align}
	0\lesssim \frac{m_{pl}}{\sqrt{N}}t\lesssim 14.
	\label{3-inflation.era0}
\end{align}
In view of (\ref{3-pure.de.sitter}) this is nothing but the region 
\begin{align}
	0\lesssim H_0t\lesssim 60.
	\label{3-inflation.era}
\end{align}
However, if we introduce an additional finite renormalization in the form of 
\begin{align*}
	\frac{\lambda}{64\pi^2}\left(-6H^2\dot{H}-2H\ddot{H}+\dot{H}^2\right), 
\end{align*}
we can relax the condition of $C_2$ by choosing appropriate $\lambda$. For instance if we choose $\lambda =25$, the solution (\ref{3-einstein.eq-begin-solution}) reads 
\begin{align}
	H(t) & \simeq H_0\left(1-C_1e^{-12.85\frac{m_{pl}}{\sqrt{N}}t}
	-C_2e^{0.1530\frac{m_{pl}}{\sqrt{N}}t}\right) \nonumber \\
	& \simeq H_0\left(1-C_1e^{-2.952H_0t}-C_2e^{0.03514H_0t}\right).
	\label{3-einstein.eq-begin-solution2}
\end{align}
$H(t)$ is depicted in Fig.~\ref{fig:energy-backreaction_inflation1}(b) for $C_1=10^{-2}$ and $C_2=10^{-2}$, which indicates that $H(t)$ is close to $H_0$ in the region (\ref{3-inflation.era0}).

\subsection{\label{3.2}End of the inflation}

In the previous subsection \ref{3.1}, we have found the inflationary solution has a sufficiently large e-folding number, and after the exponential expansion, the Hubble parameter starts to decrease. 

In this subsection we show that in the above solution the Hubble parameter becomes zero for large $t$, which indicates that the inflation will end automatically. As we have seen in the previous subsection the eq. (\ref{3-einstein.eq}) becomes invalid sometime after $H(t)$ starts to decrease. However the eq. (\ref{3-einstein.eq}) becomes valid again if $H(t)$ is sufficiently small. Therefore if we assume that $H(t)$ continues to decrease, we can use eq. (\ref{3-einstein.eq}) to describe how the inflation stops. In this region $H^4$ term on the right hand side of (\ref{3-einstein.eq}) is much smaller than $H^2$ term on the left hand side because they balanced before $H(t)$ becomes small: 
\begin{align*}
	GNH^4\ll H^2.
\end{align*}
Thus the eq. (\ref{3-einstein.eq}) reads 
\begin{align}
	& H^2=PH^2\dot{H}+QH\ddot{H}+R\dot{H}^2
	+S\left(-6H^2\dot{H}-2H\ddot{H}+\dot{H}^2\right)
	\log \frac{\sqrt{2}Hm}{\mu^2}, 
	\label{3-einstein.eq-end2}
\end{align}
where the constants $P,\> Q,\> R$ and $S$ are given by 
\begin{align}
	& P=\frac{8\pi N}{3m_{pl}^2}\times \frac{1}{2}
	\left\{\frac{1}{64\pi^2}\frac{76}{3}
	-\frac{6}{32\pi^2}\left(\log 4\pi -\gamma +\lambda \right)
	+0.01333\right\}, \nonumber \\
	& Q=\frac{8\pi N}{3m_{pl}^2}\times \frac{1}{2}
	\left\{\frac{1}{64\pi^2}\frac{20}{3}
	-\frac{2}{32\pi^2}\left(\log 4\pi -\gamma +\lambda \right)
	+0.0005572\right\}, \nonumber \\
	& R=\frac{8\pi N}{3m_{pl}^2}\times \frac{1}{2}
	\left\{-\frac{1}{64\pi^2}\frac{29}{6}
	+\frac{1}{32\pi^2}\left(\log 4\pi -\gamma +\lambda \right)
	+0.02139\right\}, \nonumber \\
	& S=-\frac{8\pi N}{3m_{pl}^2}\frac{1}{64\pi^2}.
	\label{3-einstein.eq-end2-constants}
\end{align}
Here $N$ and $\mu$ are given by (\ref{3-number.of.species}) and (\ref{3-scales}).

\begin{center}
\begin{figure}[htbp]
	\includegraphics[width=8cm]{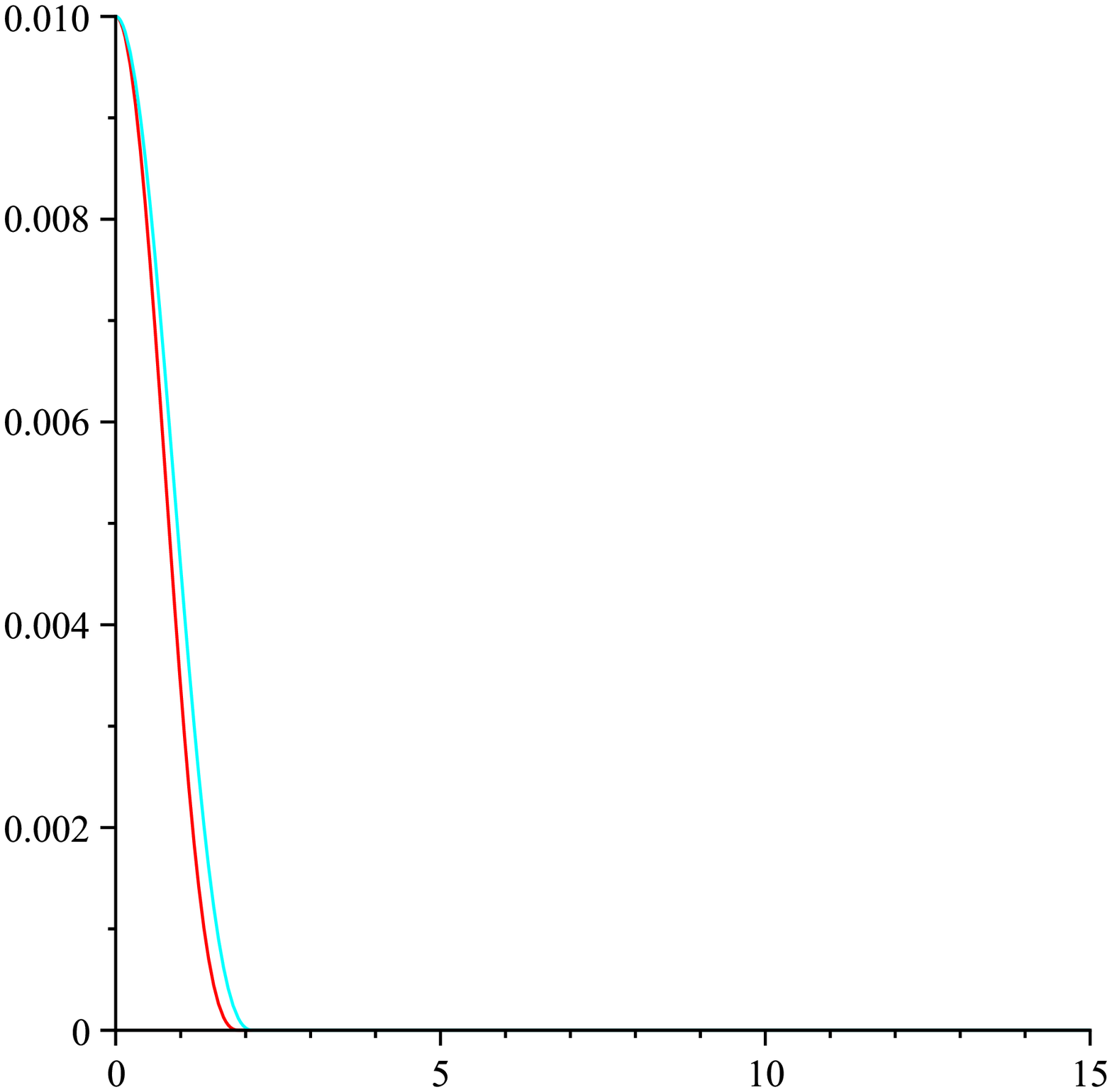}
	\includegraphics[width=8cm]{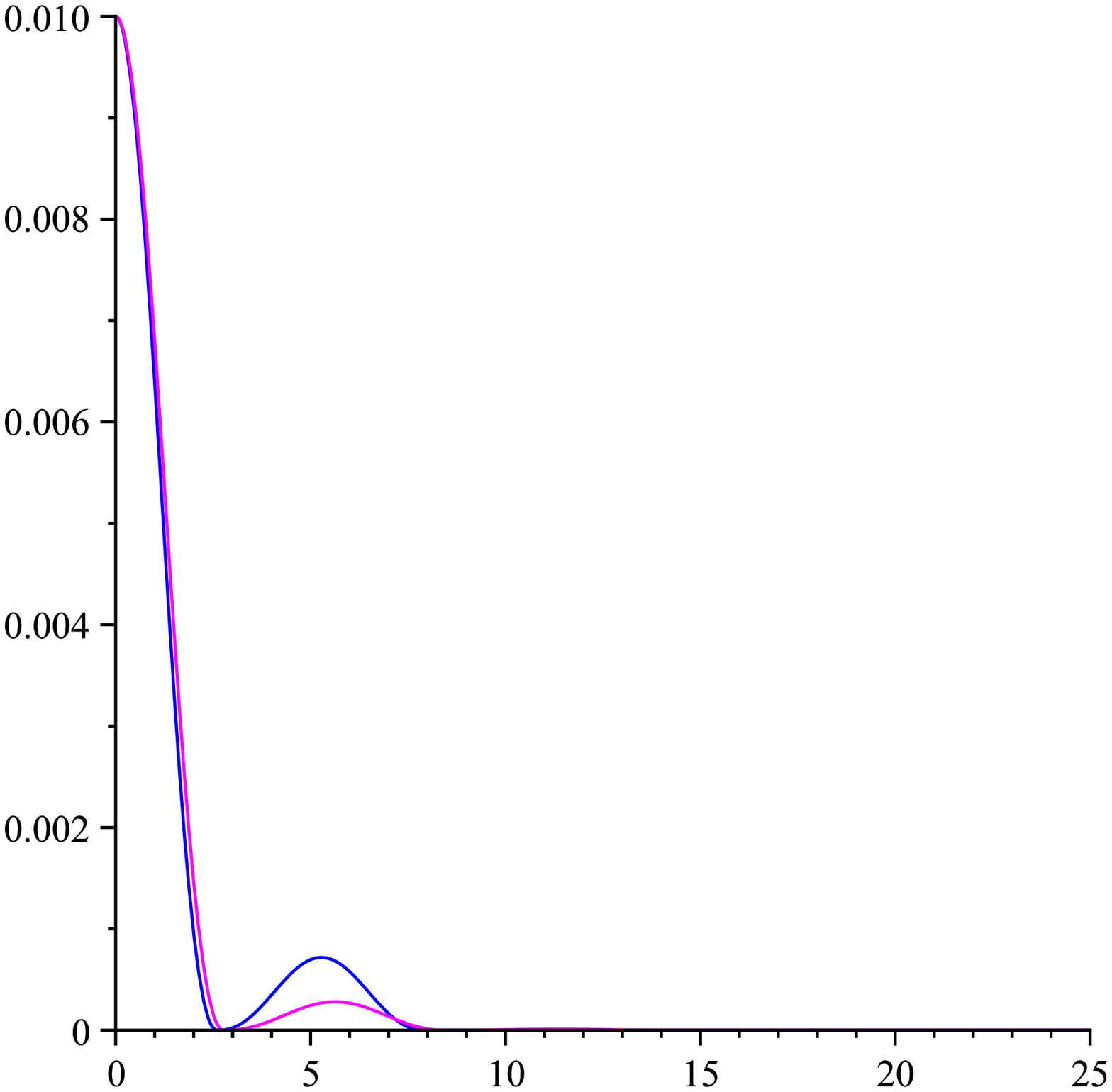}
	\put(-340,0){$\frac{m_{pl}}{\sqrt{N}}t$}
	\put(-110,0){$\frac{m_{pl}}{\sqrt{N}}t$}
	\put(-490,120){$\frac{\sqrt{N}}{m_{pl}}H(t)$}
	\thicklines
	\put(-370,140){(a)}
	\put(-350,140){$\lambda =0$}
	\put(-320,120){$m=10^{16}$ [GeV]}
	\put(-350,123){\textcolor{red}{\line(1,0){20}}}
	\put(-320,100){$m=10^{13}$ [GeV]}
	\put(-350,103){\textcolor{cyan}{\line(1,0){20}}}
	\put(-140,140){(b)}
	\put(-120,140){$\lambda =25$}
	\put(-90,120){$m=10^{16}$ [GeV]}
	\put(-120,123){\textcolor{blue}{\line(1,0){20}}}
	\put(-90,100){$m=10^{13}$ [GeV]}
	\put(-120,103){\textcolor{magenta}{\line(1,0){20}}}
	\caption{The time evolution of the Hubble parameter $H(t)$
	with the initial condition $\frac{\sqrt{N}}{m_{pl}}H(0)=10^{-2}$ 
	and $\frac{N}{m_{pl}^2}\dot{H}(0)=-10^{-6}$.}
	\label{fig:energy-backreaction_inflation2}
\end{figure}
\end{center}

The solution $H(t)$ of (\ref{3-einstein.eq-end2}) rapidly decreases as $t$ increases and then, with small oscillation, $H(t)\to 0$ as $t\to \infty$ for various choices of $m$ with $m\gg H(t)$. For example when we impose an appropriate initial condition so as to satisfy $H(0)\ll m$ and $\frac{\dot{H}}{H^2}(0)\ll 1$ and perform the numerical calculation for two cases $m=10^{16}$ and $10^{13}$ [GeV]. The results are depicted in Fig.~\ref{fig:energy-backreaction_inflation2}. Here Fig.~\ref{fig:energy-backreaction_inflation2}(a) and Fig.~\ref{fig:energy-backreaction_inflation2}(b) shows the results for $\lambda =0$ and $\lambda =25$ respectively.

\section{\label{4}Conclusion and outlook}

In the present paper we investigate the effect of the backreaction to the expansion of the space-time due to the matter fields existing in the universe. On arbitrary Friedmann-Robertson-Walker space-time, we quantize, as an example of the matter field, the scalar field and then we calculate the vacuum expectation value of the energy density. Since we are not able to obtain the exact solution of the scalar field, we make use of the Eikonal approximation to construct the solution. We then plug it in the energy density of the matter field. Combining it with the Friedmann equation we obtain a self-consistent equation that describes the time evolution of the space-time.

Thus we can determine the behavior of the early universe including the backreaction of the matter fields. As a result we find the following two eras in the early universe: 
\begin{enumerate}
\item[(1)] We obtain the inflationary solution with the e-folding number $\gtrsim 60$ if we make a fine tuning of few orders of magnitude for the initial condition. This initial condition may be determined by a scenario of the birth of the universe in string theory and/or quantum gravity.
\item[(2)] After the era (1) the Hubble parameter will continue to decrease and then with small oscillation $H(t)$ goes to zero asymptotically. Then finally the inflation ends.
\end{enumerate}
In this manner by imposing some reasonable conditions we can make a inflation scenario without inflaton. 

\begin{figure}[htbp]
	\centering
	\includegraphics[width=8cm]{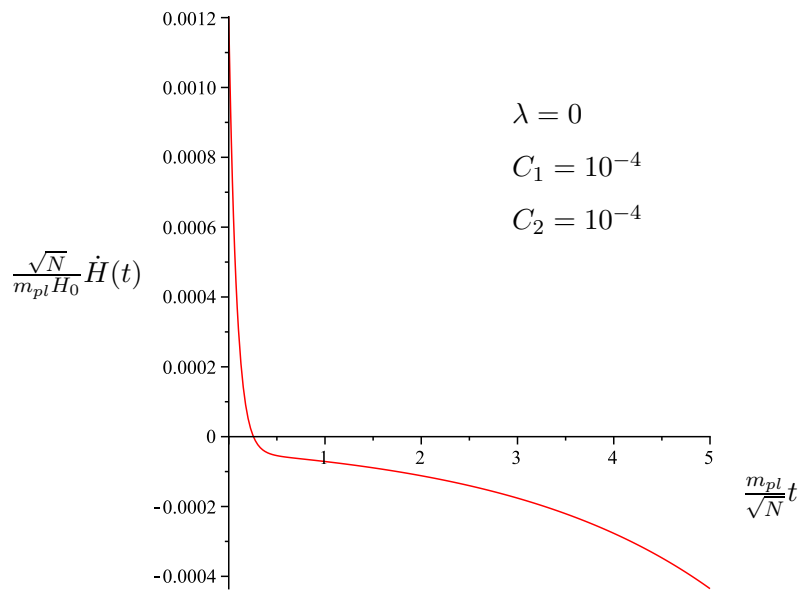}
	\put(7,37){$\frac{m_{pl}}{\sqrt{N}}t$}
	\put(-270,120){$\frac{\sqrt{N}}{m_{pl}H_0}\dot{H}(t)$}
	\put(-80,180){$\lambda =0$}
	\put(-80,160){$C_1=10^{-4}$}
	\put(-80,140){$C_2=10^{-4}$}
	\caption{$\dot{H}(t)$ 
	for the case of Fig.~\ref{fig:energy-backreaction_inflation1}(a).}
	\label{fig:energy-backreaction_inflation1-2}
\end{figure}

The problems next to be discussed are why the temperature fluctuation in the observed CMB is of the order $\frac{\delta T}{T}\sim 10^{-5}$ and very small non-Gaussianity is realized (cf. \cite{rf:planck}, $f_{NL}^{\> \text{local}}=2.7\pm 5.8$ (95\verb|%| CL)). In our previous papers \cite{rf:ours}, by using the de Sitter background without incorporating the backreaction, we estimated the CMB temperature fluctuation and non-Gaussianity $f_{NL}$ using our theory of many scalar fields, and obtained that if $\frac{\delta T}{T}\sim 10^{-5}$ then $f_{NL}<10^{-4}$. The difference between the present and previous papers ia in the fact that in the present paper we incorporated the effect of backreaction, which may not affect largely to the fluctuation, because in both cases the fluctuation is produced during the exponentially expanding era. Furthermore in the previous papers \cite{rf:ours} we had the problem that the spectral index $n_s$ exceeded 1 and contradicted the observed value. However as is mentioned in Section \ref{1} in the present paper the condition for $n_s\lesssim 1$ translated into $\dot{H}<0$ and $\left|\frac{\dot{H}}{H^2}\right|\ll 1$ is satisfied in the whole range of the inflation era (see Fig~\ref{fig:energy-backreaction_inflation1-2}).

Furthermore we consider a new mechanism to produce the temperature fluctuation in CMB that is different from our previous papers \cite{rf:ours}. In \cite{rf:ours}, we pointed out that the fluctuation of the energy density $\delta \rho =\delta T_{00}$ in the universe creates the gravitational potential $\Phi$ through the Einstein equation and turns to be the temperature fluctuation of CMB. A new mechanism we would like to discuss here is that the fluctuation $\delta t_{\text{end}}$ of the time $t_{\text{end}}$ at which the inflation ends generates the density perturbation, as in the ordinary scenario with the inflaton, $\delta =-H\delta t_{\text{end}}$. $\delta t_{\text{end}}$ can be roughly estimated as follows. First we express $t_{\text{end}}$ in terms of the initial values $H(0)$ and $\dot{H}(0)$. We define $t_{\text{end}}$ as the time $t$ when the third term of the eq. (\ref{3-einstein.eq-begin-solution1}) becomes of order $0.1$:
\begin{align}
	C_2e^{0.1012H_0t_{\text{end}}}\simeq 0.1.
	\label{4-inflation.ends}
\end{align}
On the other hand from (\ref{3-einstein.eq-begin-solution1}) we obtain 
\begin{align}
	& H(0)=H_0\left(1-C_1-C_2\right), \nonumber \\
	& \dot{H}(0)=H_0\left(2.869H_0C_1-0.1012H_0C_2\right).
	\label{4-initial.condition}
\end{align}
By combining (\ref{4-inflation.ends}) and (\ref{4-initial.condition}), we obtain 
\begin{align}
	t_{\text{end}}\simeq \frac{1}{0.1012H_0}\log \frac{0.1}{C_2}
	\simeq \frac{10}{H_0}
	\log \frac{0.1}{1-\frac{H(0)}{H_0}-\frac{\dot{H}(0)}{3H_0^2}}.
	\label{4-end_time}
\end{align}
Then $\delta t_{\text{end}}$ is given by 
\begin{align}
	\delta t_{\text{end}}\simeq -\frac{10}{H_0}
	\frac{-\frac{\delta H(0)}{H_0}-\frac{\delta \dot{H}(0)}{3H_0^2}}
	{1-\frac{H(0)}{H_0}-\frac{\dot{H}(0)}{3H_0^2}}
	\simeq 10\frac{\delta H(0)}{H_0^2}.
	\label{4-fluctuation.of.time}
\end{align}
On the other hand by taking the variation of the Friedmann equation $H_0^2\simeq \frac{8\pi G}{3}T_{00}$, we have 
\begin{align}
	H_0\delta H(0)\simeq \frac{4\pi G}{3}\delta T_{00}.
	\label{4-variation.of.Friedmann}
\end{align}
From (\ref{4-fluctuation.of.time}) and (\ref{4-variation.of.Friedmann}) we obtain the density perturbation as 
\begin{align}
	\delta \simeq -H_0\delta t_{\text{end}}
	\simeq -10\frac{\delta T_{00}}{T_{00}}\sim -\frac{10}{\sqrt{N}}.
	\label{4-density.perturbation}
\end{align}
Using the eq. (\ref{3-number.of.species}), $\delta$ is evaluated as 
\begin{align}
	\delta \sim 10^{-5}, 
	\label{4-density.perturbation-value}
\end{align}
which is consistent with the observational value. It would be interesting to investigate this new mechanism in detail.

\subsection*{Acknowledgement}
The authors acknowledge Y. Sekino for discussion in the early stage of the present work. One of the authors H.K is supported by the JSPS Grant in Aid for Scientific Research No. 22540277. M.N is also supported by the JSPS Grant in Aid for Scientific Research No. 24540293.

\appendix
\section{\label{app-airy}$\left\langle T_{00}\right\rangle$ in the region (ii) in the subsection \ref{2.1}}

In the region $k_0<k<\sqrt{V}$, there is no ultra-violet divergence, so that we can take $\epsilon =0$. The solution of the scalar field in this region is given by the eq. (\ref{2-airy.solution}), 
\begin{align}
	& \chi (t,\vec{x})=\int \frac{d^3\vec{k}}{(2\pi)^3}
	\left\{\chi_ke^{-i\vec{k}\cdot \vec{x}}a_{\vec{k}}
	+\text{c.c.}\right\}, \nonumber \\
	& \chi_k=\frac{e^{-i\frac{\pi}{4}}}{\sqrt{2}}B^{-\frac{1}{6}}\left\{
	iAi\left(B^{\frac{1}{3}}(\tau -\tau_k)\right)
	+Bi\left(B^{\frac{1}{3}}(\tau -\tau_k)\right)\right\}, \qquad 
	B\equiv V^{\prime}(\tau_k), 
	\label{app-airy.solution}
\end{align}
where the lower end of the integration of $k$ is in fact the boundary $k_0$ obtained in eq. (\ref{2-airy.boundary}), 
\begin{align*}
	k_0=\sqrt{V-\left(\frac{V^{\prime}}{2}\right)^{\frac{2}{3}}}.
\end{align*}
In order to carry out the $k$-integration of the correlation functions such as $\left\langle \chi^2\right\rangle$, let us evaluate $\tau_k$ and $B$ approximately under the conditions $\frac{\dot{H}}{H^2}\ll 1$ and $\frac{\ddot{H}}{H^3}\ll 1$. Firstly, $\tau_k$ is determined from the equation 
\begin{align*}
	k^2=V(\tau_k)
	=a^2H^2\left(2+\frac{\dot{H}}{H^2}-\frac{m^2}{H^2}\right)(\tau_k).
\end{align*}
Imposing the condition $\frac{\dot{H}}{H^2}\ll 1$, we can approximate the scale factor $a(t)$ to be the pure de Sitter one, $a\simeq -\frac{1}{H\tau}$, and obtain 
\begin{align*}
	k^2\simeq \frac{1}{\tau_k^2}
	\left(2+\frac{\dot{H}}{H^2}-\frac{m^2}{H^2}\right)
	=\frac{1}{\tau_k^2}\frac{V}{a^2H^2}.
\end{align*}
Thus $\tau_k$ is evaluated as 
\begin{align}
	\tau_k\simeq -\frac{\sqrt{V}}{aH}\frac{1}{k}.
	\label{app-tau_k.approx}
\end{align}
In the same manner $B$ becomes 
\begin{align}
	B & =V^{\prime}(\tau_k)=a^3H^3\left(
	4+6\frac{\dot{H}}{H^2}+\frac{\ddot{H}}{H^3}-\frac{m^2}{H^2}
	\right)(\tau_k)\simeq -\frac{1}{\tau_k^3}\frac{V^{\prime}}{a^3H^3} 
	\nonumber \\
	& \simeq \frac{V^{\prime}}{V^{\frac{3}{2}}}k^3.
	\label{app-B.approx}
\end{align}

Then we would like to calculate the four kinds of correlation functions in eq. (\ref{2-energy-momentum.tensor}) in order. Firstly, $\left\langle \chi^2\right\rangle$ is calculated as 
\begin{align*}
	\left\langle \chi^2\right\rangle & =\int \frac{d^3\vec{k}}{(2\pi )^3}
	\left|\chi_k\right|^2 \\
	& =\frac{1}{(2\pi )^3}
	\frac{2\pi^{\frac{3}{2}}}{\Gamma \left(\frac{3}{2}\right)}
	\int_{k_0}^{\sqrt{V}}dk\> k^2\times \frac{1}{2B^{\frac{1}{3}}}
	\left\{Ai^2\left(B^{\frac{1}{3}}(\tau -\tau_k)\right)
	+Bi^2\left(B^{\frac{1}{3}}(\tau -\tau_k)\right)\right\} \\
	& =\frac{1}{4\pi^2}\frac{\sqrt{V}}{(V^{\prime})^{\frac{1}{3}}}
	\int_{k_0}^{\sqrt{V}}dk\> k
	\left\{Ai^2\left(\frac{(V^{\prime})^{\frac{1}{3}}}{\sqrt{V}}
	\frac{\sqrt{V}-k}{aH}\right)
	+Bi^2\left(\frac{(V^{\prime})^{\frac{1}{3}}}{\sqrt{V}}
	\frac{\sqrt{V}-k}{aH}\right)\right\}.
\end{align*}
Here (\ref{app-tau_k.approx}) and (\ref{app-B.approx}) are used to obtain the last line. After the change of variable $x=\frac{(V^{\prime})^{\frac{1}{3}}}{\sqrt{V}}\frac{\sqrt{V}-k}{aH}$, this integration turns to be 
\begin{align}
	\left\langle \chi^2\right\rangle 
	& =\frac{aH}{4\pi^2}\frac{V^{\frac{3}{2}}}{(V^{\prime})^{\frac{2}{3}}}
	\int_0^{x_0}dx\> \left(1-\frac{aH}{(V^{\prime})^{\frac{1}{3}}}x\right)
	\left\{Ai^2(x)+Bi^2(x)\right\} \nonumber \\
	& =\frac{a^2H^2}{4\pi^2}
	\left(2+\frac{\dot{H}}{H^2}-\frac{m^2}{H^2}\right)^{\frac{3}{2}}
	\left(4+6\frac{\dot{H}}{H^2}+\frac{\ddot{H}}{H^3}-2\frac{m^2}{H^2}
	\right)^{-\frac{2}{3}}
	\int_0^{x_0}dx\> \left\{Ai^2(x)+Bi^2(x)\right\} \nonumber \\
	& \quad -\frac{a^2H^2}{4\pi^2}
	\left(2+\frac{\dot{H}}{H^2}-\frac{m^2}{H^2}\right)^{\frac{3}{2}}
	\left(4+6\frac{\dot{H}}{H^2}+\frac{\ddot{H}}{H^3}-2\frac{m^2}{H^2}
	\right)^{-1}
	\int_0^{x_0}dx\> x\left\{Ai^2(x)+Bi^2(x)\right\}.
	\label{app-correlation-mass-airy}
\end{align}
Here we define the upper end of the integration region as 
\begin{align*}
	x_0 & \equiv \frac{(V^{\prime})^{\frac{1}{3}}}{\sqrt{V}}
	\frac{\sqrt{V}-k_0}{aH} \\
	& =\frac{(V^{\prime})^{\frac{1}{3}}}{aH}
	\left\{1-\sqrt{1-\frac{1}{V}\left(\frac{V^{\prime}}{2}\right)
	^{\frac{2}{3}}}\right\}.
\end{align*}
By assuming that $\frac{\dot{H}}{H^2},\> \frac{\ddot{H}}{H^3}$ and $\frac{m^2}{H^2}$ are small, this upper end can be approximated as 
\begin{align}
	x_0 & =\frac{(V^{\prime})^{\frac{1}{3}}}{aH}
	\left\{1-\sqrt{1-\frac{1}{V}\left(\frac{V^{\prime}}{2}
	\right)^{\frac{2}{3}}}\right\} \nonumber \\
	& =\left(4+6\frac{\dot{H}}{H^2}+\frac{\ddot{H}}{H^3}-2\frac{m^2}{H^2}
	\right)^{\frac{1}{3}}\left\{1-\sqrt{1-\frac{1}{2+\frac{\dot{H}}{H^2}
	-\frac{m^2}{H^2}}
	\left(\frac{4+6\frac{\dot{H}}{H^2}+\frac{\ddot{H}}{H^3}
	-2\frac{m^2}{H^2}}{2}\right)^{\frac{2}{3}}}\right\} \nonumber \\
	& \simeq 4^{\frac{1}{3}}\left\{1-\sqrt{1-\frac{2^{\frac{2}{3}}}{2}}
	\right\}.
	\label{app-x.integral-top}
\end{align}
Then we perform the integrations including the Airy functions numerically and expand them in terms of $\frac{\dot{H}}{H^2},\> \frac{\ddot{H}}{H^3}$ and $\frac{m^2}{H^2}$.

In the same way, we calculate the correlations $\left\langle \left(\nabla \chi \right)^2\right\rangle$, $\left\langle \left(\chi^{\prime}\right)^2\right\rangle$ and $\left\langle \chi^{\prime}\chi +\chi \chi^{\prime}\right\rangle$ as follows: 
\begin{align}
	\left\langle \left(\nabla \chi \right)^2\right\rangle 
	& =\int \frac{d^3\vec{k}}{(2\pi )^3}k^2\left|\chi_k\right|^2 
	\nonumber \\
	& =\frac{1}{(2\pi )^3}
	\frac{2\pi^{\frac{3}{2}}}{\Gamma \left(\frac{3}{2}\right)}
	\int_{k_0}^{\sqrt{V}}dk\> k^4\times \frac{1}{2B^{\frac{1}{3}}}
	\left\{Ai^2\left(B^{\frac{1}{3}}(\tau -\tau_k)\right)
	+Bi^2\left(B^{\frac{1}{3}}(\tau -\tau_k)\right)\right\} \nonumber \\
	& =\frac{1}{4\pi^2}\frac{\sqrt{V}}{(V^{\prime})^{\frac{1}{3}}}
	\int_{k_0}^{\sqrt{V}}dk\> k^3
	\left\{Ai^2\left(\frac{(V^{\prime})^{\frac{1}{3}}}{\sqrt{V}}
	\frac{\sqrt{V}-k}{aH}\right)
	+Bi^2\left(\frac{(V^{\prime})^{\frac{1}{3}}}{\sqrt{V}}
	\frac{\sqrt{V}-k}{aH}\right)\right\} \nonumber \\
	& =\frac{aH}{4\pi^2}\frac{V^{\frac{5}{2}}}{(V^{\prime})^{\frac{2}{3}}}
	\int_0^{x_0}dx\> 
	\left(1-\frac{aH}{(V^{\prime})^{\frac{1}{3}}}x\right)^3
	\left\{Ai^2(x)+Bi^2(x)\right\} \nonumber \\
	& =\frac{a^4H^4}{4\pi^2}
	\left(2+\frac{\dot{H}}{H^2}-\frac{m^2}{H^2}\right)^{\frac{5}{2}}
	\left(4+6\frac{\dot{H}}{H^2}+\frac{\ddot{H}}{H^3}-2\frac{m^2}{H^2}
	\right)^{-\frac{2}{3}}
	\int_0^{x_0}dx\> \left\{Ai^2(x)+Bi^2(x)\right\} \nonumber \\
	& \quad -\frac{a^4H^4}{4\pi^2}
	\left(2+\frac{\dot{H}}{H^2}-\frac{m^2}{H^2}\right)^{\frac{5}{2}}
	\left(4+6\frac{\dot{H}}{H^2}+\frac{\ddot{H}}{H^3}-2\frac{m^2}{H^2}
	\right)^{-1}
	\times 3\int_0^{x_0}dx\> x\left\{Ai^2(x)+Bi^2(x)\right\} \nonumber \\
	& \quad +\frac{a^4H^4}{4\pi^2}
	\left(2+\frac{\dot{H}}{H^2}-\frac{m^2}{H^2}\right)^{\frac{5}{2}}
	\left(4+6\frac{\dot{H}}{H^2}+\frac{\ddot{H}}{H^3}-2\frac{m^2}{H^2}
	\right)^{-\frac{4}{3}}
	\times 3\int_0^{x_0}dx\> x^2\left\{Ai^2(x)+Bi^2(x)\right\} \nonumber \\
	& \quad -\frac{a^4H^4}{4\pi^2}
	\left(2+\frac{\dot{H}}{H^2}-\frac{m^2}{H^2}\right)^{\frac{5}{2}}
	\left(4+6\frac{\dot{H}}{H^2}+\frac{\ddot{H}}{H^3}-2\frac{m^2}{H^2}
	\right)^{-\frac{5}{3}}
	\int_0^{x_0}dx\> x^3\left\{Ai^2(x)+Bi^2(x)\right\}.
	\label{app-correlation-nabla-airy}
\end{align}
\begin{align}
	\left\langle \left(\chi^{\prime}\right)^2\right\rangle 
	& =\int \frac{d^3\vec{k}}{(2\pi )^3}\left|\chi_k^{\prime}\right|^2 
	\nonumber \\
	& =\frac{1}{(2\pi )^3}
	\frac{2\pi^{\frac{3}{2}}}{\Gamma \left(\frac{3}{2}\right)}
	\int_{k_0}^{\sqrt{V}}dk\> k^2\times \frac{B^{\frac{1}{3}}}{2}
	\left\{Ai^{\prime 2}\left(B^{\frac{1}{3}}(\tau -\tau_k)\right)
	+Bi^{\prime 2}\left(B^{\frac{1}{3}}(\tau -\tau_k)\right)\right\} 
	\nonumber \\
	& =\frac{1}{4\pi^2}\frac{(V^{\prime})^{\frac{1}{3}}}{\sqrt{V}}
	\int_{k_0}^{\sqrt{V}}dk\> k^3
	\left\{Ai^{\prime 2}\left(\frac{(V^{\prime})^{\frac{1}{3}}}{\sqrt{V}}
	\frac{\sqrt{V}-k}{aH}\right)
	+Bi^{\prime 2}\left(\frac{(V^{\prime})^{\frac{1}{3}}}{\sqrt{V}}
	\frac{\sqrt{V}-k}{aH}\right)\right\} \nonumber \\
	& =\frac{aH}{4\pi^2}V^{\frac{3}{2}}
	\int_0^{x_0}dx\> 
	\left(1-\frac{aH}{(V^{\prime})^{\frac{1}{3}}}x\right)^3
	\left\{Ai^{\prime 2}(x)+Bi^{\prime 2}(x)\right\} \nonumber \\
	& =\frac{a^4H^4}{4\pi^2}
	\left(2+\frac{\dot{H}}{H^2}-\frac{m^2}{H^2}\right)^{\frac{3}{2}}
	\int_0^{x_0}dx\> \left\{Ai^{\prime 2}(x)+Bi^{\prime 2}(x)\right\} 
	\nonumber \\
	& \quad -\frac{a^4H^4}{4\pi^2}
	\left(2+\frac{\dot{H}}{H^2}-\frac{m^2}{H^2}\right)^{\frac{3}{2}}
	\left(4+6\frac{\dot{H}}{H^2}+\frac{\ddot{H}}{H^3}-2\frac{m^2}{H^2}
	\right)^{-\frac{1}{3}}
	\times 3\int_0^{x_0}dx\> 
	x\left\{Ai^{\prime 2}(x)+Bi^{\prime 2}(x)\right\} \nonumber \\
	& \quad +\frac{a^4H^4}{4\pi^2}
	\left(2+\frac{\dot{H}}{H^2}-\frac{m^2}{H^2}\right)^{\frac{3}{2}}
	\left(4+6\frac{\dot{H}}{H^2}+\frac{\ddot{H}}{H^3}-2\frac{m^2}{H^2}
	\right)^{-\frac{2}{3}}
	\times 3\int_0^{x_0}dx\> 
	x^2\left\{Ai^{\prime 2}(x)+Bi^{\prime 2}(x)\right\} \nonumber \\
	& \quad -\frac{a^4H^4}{4\pi^2}
	\left(2+\frac{\dot{H}}{H^2}-\frac{m^2}{H^2}\right)^{\frac{3}{2}}
	\left(4+6\frac{\dot{H}}{H^2}+\frac{\ddot{H}}{H^3}-2\frac{m^2}{H^2}
	\right)^{-1}
	\int_0^{x_0}dx\> x^3\left\{Ai^{\prime 2}(x)+Bi^{\prime 2}(x)\right\}.
	\label{app-correlation-kinetic-airy}
\end{align}
\begin{align}
	& \left\langle \chi^{\prime}\chi +\chi^{\prime}\chi \right\rangle 
	=\int \frac{d^3\vec{k}}{(2\pi )^3}
	\left(\chi_k^{\prime}\chi_k^{\ast}+\chi_k^{\prime \ast}\chi_k \right) 
	\nonumber \\
	& =\frac{1}{(2\pi )^3}
	\frac{2\pi^{\frac{3}{2}}}{\Gamma \left(\frac{3}{2}\right)}
	\int_{k_0}^{\sqrt{V}}dk\> k^2
	\left\{Ai^{\prime}\left(B^{\frac{1}{3}}(\tau -\tau_k)\right)
	Ai\left(B^{\frac{1}{3}}(\tau -\tau_k)\right)
	+Bi^{\prime}\left(B^{\frac{1}{3}}(\tau -\tau_k)\right)
	Bi\left(B^{\frac{1}{3}}(\tau -\tau_k)\right)\right\} 
	\nonumber \\
	& =\frac{1}{2\pi^2}\int_{k_0}^{\sqrt{V}}dk\> k^2 \nonumber \\
	& \qquad \qquad \quad \times 
	\left\{Ai^{\prime}\left(\frac{(V^{\prime})^{\frac{1}{3}}}{\sqrt{V}}
	\frac{\sqrt{V}-k}{aH}\right)
	Ai\left(\frac{(V^{\prime})^{\frac{1}{3}}}{\sqrt{V}}
	\frac{\sqrt{V}-k}{aH}\right)
	+Bi^{\prime}\left(\frac{(V^{\prime})^{\frac{1}{3}}}{\sqrt{V}}
	\frac{\sqrt{V}-k}{aH}\right)
	Bi\left(\frac{(V^{\prime})^{\frac{1}{3}}}{\sqrt{V}}
	\frac{\sqrt{V}-k}{aH}\right)\right\} \nonumber \\
	& =\frac{aH}{2\pi^2}\frac{V^{\frac{3}{2}}}{(V^{\prime})^{\frac{1}{3}}}
	\int_0^{x_0}dx\> 
	\left(1-\frac{aH}{(V^{\prime})^{\frac{1}{3}}}x\right)^2
	\left\{Ai^{\prime}(x)Ai(x)+Bi^{\prime}(x)Bi(x)\right\} \nonumber \\
	& =\frac{a^3H^3}{2\pi^2}
	\left(2+\frac{\dot{H}}{H^2}-\frac{m^2}{H^2}\right)^{\frac{3}{2}}
	\left(4+6\frac{\dot{H}}{H^2}+\frac{\ddot{H}}{H^3}-2\frac{m^2}{H^2}
	\right)^{-\frac{1}{3}}
	\int_0^{x_0}dx\> 
	\left\{Ai^{\prime}(x)Ai(x)+Bi^{\prime}(x)Bi(x)\right\} \nonumber \\
	& \quad -\frac{a^3H^3}{2\pi^2}
	\left(2+\frac{\dot{H}}{H^2}-\frac{m^2}{H^2}\right)^{\frac{3}{2}}
	\left(4+6\frac{\dot{H}}{H^2}+\frac{\ddot{H}}{H^3}-2\frac{m^2}{H^2}
	\right)^{-\frac{2}{3}}
	\times 2\int_0^{x_0}dx\> 
	x\left\{Ai^{\prime}(x)Ai(x)+Bi^{\prime}(x)Bi(x)\right\} \nonumber \\
	& \quad +\frac{a^3H^3}{2\pi^2}
	\left(2+\frac{\dot{H}}{H^2}-\frac{m^2}{H^2}\right)^{\frac{3}{2}}
	\left(4+6\frac{\dot{H}}{H^2}+\frac{\ddot{H}}{H^3}-2\frac{m^2}{H^2}
	\right)^{-1}
	\int_0^{x_0}dx\> 
	x^2\left\{Ai^{\prime}(x)Ai(x)+Bi^{\prime}(x)Bi(x)\right\}.
	\label{app-correlation-cross-airy}
\end{align}

\section{\label{app-damping}$\left\langle T_{00}\right\rangle$ in the region (iii) in the subsection \ref{2.1}}

Similarly to the region (ii), we do not have the ultra-violet divergence in the region $0\leq k<k_0$, so that we take $\epsilon =0$. As we stated in the subsection \ref{2.1}, from (\ref{2-airy.boundary}), we have the upper end of this region 
\begin{align*}
	k_0=\sqrt{V-\left(\frac{V^{\prime}}{2}\right)^{\frac{2}{3}}}, 
\end{align*}
and the solution of the scalar field is given by eq. (\ref{2-damping.solution}): 
\begin{align}
	& \chi (t,\vec{x})=\int \frac{d^3\vec{k}}{(2\pi)^3}
	\left\{\chi_ke^{-i\vec{k}\cdot \vec{x}}a_{\vec{k}}
	+\text{c.c.}\right\}, \nonumber \\
	& \chi_k=\frac{e^{-i\frac{\pi}{4}}}{\sqrt{2}}
	\frac{1}{\sqrt[4]{V(\tau )-k^2}}\left\{
	\frac{i}{2}e^{-\int_{\tau_k}^{\tau}d\tau^{\prime}
	\sqrt{V(\tau^{\prime})-k^2}}
	+e^{+\int_{\tau_k}^{\tau}d\tau^{\prime}
	\sqrt{V(\tau^{\prime})-k^2}}\right\}.
	\label{app-damping.solution}
\end{align}
Here let us calculate $\tau$ integration of the exponent of the Fourier mode $\chi_k$
\begin{align*}
	\int_{\tau_k}^{\tau}d\tau^{\prime}\sqrt{V(\tau^{\prime})-k^2}
	& =\int_{\tau_k}^{\tau}d\tau^{\prime}
	\left\{a^2\left(2H^2+\dot{H}-m^2\right)-k^2\right\}^{\frac{1}{2}}, 
\end{align*}
under the condition $\frac{\dot{H}}{H^2}\ll 1$. In this situation we can approximate the scale factor as the pure de Sitter solution $a(t)\simeq -\frac{1}{H\tau}$, and we obtain 
\begin{align}
	\int_{\tau_k}^{\tau} & d\tau^{\prime}\sqrt{V(\tau^{\prime})-k^2}
	\simeq -\int_{\tau_k}^{\tau}\frac{d\tau^{\prime}}{\tau^{\prime}}
	\sqrt{2+\frac{\dot{H}}{H^2}-\frac{m^2}{H^2}-k^2\tau^{\prime 2}} 
	\nonumber \\
	& \simeq -\sqrt{2+\frac{\dot{H}}{H^2}-\frac{m^2}{H^2}-k^2\tau^2}
	+\frac{1}{2}\sqrt{2+\frac{\dot{H}}{H^2}-\frac{m^2}{H^2}}
	\log \frac{\sqrt{2+\frac{\dot{H}}{H^2}-\frac{m^2}{H^2}}
	+\sqrt{2+\frac{\dot{H}}{H^2}-\frac{m^2}{H^2}-k^2\tau^2}}
	{\sqrt{2+\frac{\dot{H}}{H^2}-\frac{m^2}{H^2}}
	-\sqrt{2+\frac{\dot{H}}{H^2}-\frac{m^2}{H^2}-k^2\tau^2}} \nonumber \\
	& =-\frac{\sqrt{V-k^2}}{aH}+\frac{1}{2}\frac{\sqrt{V}}{aH}
	\log \frac{\sqrt{V}+\sqrt{V-k^2}}{\sqrt{V}-\sqrt{V-k^2}}.
	\label{app-damping.exponent}
\end{align}
Now we evaluate the four kinds of correlation functions included in the energy density (\ref{2-energy-momentum.tensor}). Firstly, as for the correlation $\left\langle \chi^2\right\rangle$, applying the result (\ref{app-damping.exponent}), we find 
\begin{align*}
	\left\langle \chi^2\right\rangle & =\int \frac{d^3\vec{k}}{(2\pi )^3}
	\left|\chi_k\right|^2 \\
	& =\frac{1}{(2\pi )^3}
	\frac{2\pi^{\frac{3}{2}}}{\Gamma \left(\frac{3}{2}\right)}
	\int_0^{k_0}dk\> k^2\frac{1}{2\sqrt{V-k^2}}\left(
	\frac{1}{4}e^{-2\int_{\tau_k}^{\tau}d\tau^{\prime}\sqrt{V-k^2}}
	+e^{+2\int_{\tau_k}^{\tau}d\tau^{\prime}\sqrt{V-k^2}}\right) \\
	& =\frac{1}{4\pi^2}\int_0^{k_0}dk\> \frac{k^2}{\sqrt{V-k^2}}
	\left\{\frac{1}{4}\left(\textstyle 
	\frac{\sqrt{V}+\sqrt{V-k^2}}{\sqrt{V}-\sqrt{V-k^2}}
	\right)^{-\frac{\sqrt{V}}{aH}}e^{2\frac{\sqrt{V-k^2}}{aH}}
	+\left(\textstyle 
	\frac{\sqrt{V}+\sqrt{V-k^2}}{\sqrt{V}-\sqrt{V-k^2}}
	\right)^{\frac{\sqrt{V}}{aH}}e^{-2\frac{\sqrt{V-k^2}}{aH}}\right\}.
\end{align*}
By changing the integration variable as $z=\frac{\sqrt{V-k^2}}{\sqrt{V}}$, and using the lower end of $z$ integration, 
\begin{align*}
	z_0\equiv \frac{\sqrt{V-k_0^2}}{\sqrt{V}}
	=\frac{1}{\sqrt{V}}\left(\frac{V^{\prime}}{2}\right)^{\frac{1}{3}}, 
\end{align*}
we rewrite $\left\langle \chi^2\right\rangle$ as 
\begin{align*}
	\left\langle \chi^2\right\rangle & =\frac{V}{4\pi^2}
	\int_{z_0}^1dz\> \sqrt{1-z^2}\left\{\frac{1}{4}
	\left(\frac{1+z}{1-z}\right)^{-\frac{\sqrt{V}}{aH}}
	e^{2\frac{\sqrt{V}}{aH}z}
	+\left(\frac{1+z}{1-z}\right)^{\frac{\sqrt{V}}{aH}}
	e^{-2\frac{\sqrt{V}}{aH}z}\right\}.
\end{align*}
Furthermore we expand $\frac{\sqrt{V}}{aH}=\sqrt{2+\frac{\dot{H}}{H^2}-\frac{m^2}{H^2}}$ with respect to $\frac{\dot{H}}{H^2}$ and $\frac{m^2}{H^2}$: 
\begin{align}
	& \left\langle \chi^2\right\rangle \nonumber \\
	& =\frac{a^2H^2}{4\pi^2}
	\left(2+\frac{\dot{H}}{H^2}-\frac{m^2}{H^2}\right) \nonumber \\
	& \quad \times \Bigg[
	\int_{z_0}^1dz\> \sqrt{1-z^2}\bigg\{\textstyle \frac{1}{4}
	\left(\frac{1+z}{1-z}\right)^{-\sqrt{2}}e^{2\sqrt{2}z}
	+\left(\frac{1+z}{1-z}\right)^{\sqrt{2}}e^{-2\sqrt{2}z}\bigg\} 
	\nonumber \\
	& \qquad \quad 
	-\frac{\sqrt{2}}{4}\left(\frac{\dot{H}}{H^2}-\frac{m^2}{H^2}\right) 
	\int_{z_0}^1dz\> \sqrt{1-z^2}\textstyle 
	\left(\log \frac{1+z}{1-z}-2z\right)\bigg\{\frac{1}{4}
	\left(\frac{1+z}{1-z}\right)^{-\sqrt{2}}e^{2\sqrt{2}z}
	-\left(\frac{1+z}{1-z}\right)^{\sqrt{2}}e^{-2\sqrt{2}z}\bigg\} 
	\nonumber \\
	& \qquad \quad 
	+\frac{1}{32}\left(\frac{\dot{H}}{H^2}-\frac{m^2}{H^2}\right)^2 
	\nonumber \\
	& \qquad \qquad \times 
	\int_{z_0}^1dz\> \sqrt{1-z^2}\textstyle 
	\bigg\{\frac{1}{4}\left((\sqrt{2}-8z)\log \frac{1+z}{1-z}
	+2\left(\log \frac{1+z}{1-z}\right)^2-2\sqrt{2}z+8z^2\right)
	\left(\frac{1+z}{1-z}\right)^{-\sqrt{2}}e^{2\sqrt{2}z} \nonumber \\
	& \qquad \qquad \qquad \qquad \qquad \qquad \textstyle 
	-\left((\sqrt{2}+8z)\log \frac{1+z}{1-z}
	-2\left(\log \frac{1+z}{1-z}\right)^2-2\sqrt{2}z-8z^2\right)
	\left(\frac{1+z}{1-z}\right)^{\sqrt{2}}e^{-2\sqrt{2}z}\bigg\} 
	\nonumber \\
	& \qquad \quad +\cdots \Bigg].
	\label{app-correlation-mass}
\end{align}
Here, in each terms with $\left(\frac{\dot{H}}{H^2}-\frac{m^2}{H^2}\right)^n,\> (n=0,1,2,\cdots )$, we approximate the lower end $z_0$ of the $z$ integration as 
\begin{align}
	z_0=\frac{1}{\sqrt{V}}\left(\frac{V^{\prime}}{2}\right)^{\frac{1}{3}}
	=\frac{1}{\sqrt{2+\frac{\dot{H}}{H^2}-\frac{m^2}{H^2}}}
	\left(\frac{4+6\frac{\dot{H}}{H^2}+\frac{\ddot{H}}{H^3}
	-2\frac{m^2}{H^2}}{2}\right)^{\frac{1}{3}}
	\simeq \frac{2^{\frac{1}{3}}}{\sqrt{2}}.
	\label{app-z.integral-bottom}
\end{align}
Here we have used (\ref{2-potential}) and (\ref{2-potential-prime}) to obtain the second equality. In practice we calculate the $z$ integrations numerically.

In the same manner we evaluate the rest correlations $\left\langle \left(\nabla \chi \right)^2\right\rangle$, $\left\langle \left(\chi^{\prime}\right)^2\right\rangle$ and $\left\langle \chi^{\prime}\chi +\chi \chi^{\prime}\right\rangle$.
\begin{align}
	& \left\langle \left(\nabla \chi \right)^2\right\rangle 
	=\int \frac{d^3\vec{k}}{(2\pi )^3}k^2\left|\chi_k\right|^2 \nonumber \\
	& =\frac{1}{(2\pi )^3}
	\frac{2\pi^{\frac{3}{2}}}{\Gamma \left(\frac{3}{2}\right)}
	\int_0^{k_0}dk\> k^4\frac{1}{2\sqrt{V-k^2}}\left(
	\frac{1}{4}e^{-2\int_{\tau_k}^{\tau}d\tau^{\prime}\sqrt{V-k^2}}
	+e^{+2\int_{\tau_k}^{\tau}d\tau^{\prime}\sqrt{V-k^2}}\right) 
	\nonumber \\
	& =\frac{1}{4\pi^2}\int_0^{k_0}dk\> \frac{k^4}{\sqrt{V-k^2}}
	\left\{\frac{1}{4}\left(\textstyle 
	\frac{\sqrt{V}+\sqrt{V-k^2}}{\sqrt{V}-\sqrt{V-k^2}}
	\right)^{-\frac{\sqrt{V}}{aH}}e^{2\frac{\sqrt{V-k^2}}{aH}}
	+\left(\textstyle 
	\frac{\sqrt{V}+\sqrt{V-k^2}}{\sqrt{V}-\sqrt{V-k^2}}
	\right)^{\frac{\sqrt{V}}{aH}}e^{-2\frac{\sqrt{V-k^2}}{aH}}\right\} 
	\nonumber \\
	& =\frac{V^2}{4\pi^2}
	\int_{z_0}^1dz\> \left(1-z^2\right)^{\frac{3}{2}}\left\{\frac{1}{4}
	\left(\frac{1+z}{1-z}\right)^{-\frac{\sqrt{V}}{aH}}
	e^{2\frac{\sqrt{V}}{aH}z}
	+\left(\frac{1+z}{1-z}\right)^{\frac{\sqrt{V}}{aH}}
	e^{-2\frac{\sqrt{V}}{aH}z}\right\} \nonumber \\
	& =\frac{a^4H^4}{4\pi^2}
	\left(2+\frac{\dot{H}}{H^2}-\frac{m^2}{H^2}\right)^2 \nonumber \\
	& \quad \times \Bigg[
	\int_{z_0}^1dz\> \left(1-z^2\right)^{\frac{3}{2}}
	\bigg\{\textstyle \frac{1}{4}
	\left(\frac{1+z}{1-z}\right)^{-\sqrt{2}}e^{2\sqrt{2}z}
	+\left(\frac{1+z}{1-z}\right)^{\sqrt{2}}e^{-2\sqrt{2}z}\bigg\} 
	\nonumber \\
	& \qquad \quad 
	-\frac{\sqrt{2}}{4}\left(\frac{\dot{H}}{H^2}-\frac{m^2}{H^2}\right) 
	\int_{z_0}^1dz\> \left(1-z^2\right)^{\frac{3}{2}}\textstyle 
	\left(\log \frac{1+z}{1-z}-2z\right)\bigg\{\frac{1}{4}
	\left(\frac{1+z}{1-z}\right)^{-\sqrt{2}}e^{2\sqrt{2}z}
	-\left(\frac{1+z}{1-z}\right)^{\sqrt{2}}e^{-2\sqrt{2}z}\bigg\} 
	\nonumber \\
	& \qquad \quad 
	+\frac{1}{32}\left(\frac{\dot{H}}{H^2}-\frac{m^2}{H^2}\right)^2 
	\nonumber \\
	& \qquad \qquad \times 
	\int_{z_0}^1dz\> \left(1-z^2\right)^{\frac{3}{2}}\textstyle 
	\bigg\{\frac{1}{4}\left((\sqrt{2}-8z)\log \frac{1+z}{1-z}
	+2\left(\log \frac{1+z}{1-z}\right)^2-2\sqrt{2}z+8z^2\right)
	\left(\frac{1+z}{1-z}\right)^{-\sqrt{2}}e^{2\sqrt{2}z} \nonumber \\
	& \qquad \qquad \qquad \qquad \qquad \qquad \quad \textstyle 
	-\left((\sqrt{2}+8z)\log \frac{1+z}{1-z}
	-2\left(\log \frac{1+z}{1-z}\right)^2-2\sqrt{2}z-8z^2\right)
	\left(\frac{1+z}{1-z}\right)^{\sqrt{2}}e^{-2\sqrt{2}z}\bigg\} 
	\nonumber \\
	& \qquad \quad +\cdots \Bigg].
	\label{app-correlation-nabla}
\end{align}
\begin{align}
	& \left\langle \left(\chi^{\prime}\right)^2\right\rangle 
	=\int \frac{d^3\vec{k}}{(2\pi )^3}\left|\chi_k^{\prime}\right|^2 
	\nonumber \\
	& =\frac{1}{(2\pi )^3}
	\frac{2\pi^{\frac{3}{2}}}{\Gamma \left(\frac{3}{2}\right)}
	\int_0^{k_0}dk\> k^2\frac{1}{2}\Bigg[
	\frac{1}{4}\left\{\frac{V^{\prime}}{4}(V-k^2)^{-\frac{5}{4}}
	+(V-k^2)^{\frac{1}{4}}\right\}^2
	e^{-2\int_{\tau_k}^{\tau}d\tau^{\prime}\sqrt{V-k^2}} \nonumber \\
	& \qquad \qquad \qquad \qquad \qquad \qquad 
	+\left\{\frac{V^{\prime}}{4}(V-k^2)^{-\frac{5}{4}}
	-(V-k^2)^{\frac{1}{4}}\right\}^2
	e^{+2\int_{\tau_k}^{\tau}d\tau^{\prime}\sqrt{V-k^2}}\Bigg] 
	\nonumber \\
	& =\frac{1}{4\pi^2}\int_0^{k_0}dk\> k^2
	\Bigg[\frac{1}{4}\left\{\frac{V^{\prime}}{4}(V-k^2)^{-\frac{5}{4}}
	+(V-k^2)^{\frac{1}{4}}\right\}^2
	\left(\textstyle \frac{\sqrt{V}+\sqrt{V-k^2}}{\sqrt{V}-\sqrt{V-k^2}}
	\right)^{-\frac{\sqrt{V}}{aH}}e^{2\frac{\sqrt{V-k^2}}{aH}} \nonumber \\
	& \qquad \qquad \qquad \qquad 
	+\left\{\frac{V^{\prime}}{4}(V-k^2)^{-\frac{5}{4}}
	-(V-k^2)^{\frac{1}{4}}\right\}^2
	\left(\textstyle \frac{\sqrt{V}+\sqrt{V-k^2}}{\sqrt{V}-\sqrt{V-k^2}}
	\right)^{\frac{\sqrt{V}}{aH}}e^{-2\frac{\sqrt{V-k^2}}{aH}}\Bigg] 
	\nonumber \\
	& =\frac{V^2}{4\pi^2}
	\int_{z_0}^1dz\> z^2\sqrt{1-z^2}\left[\frac{1}{4}
	\left\{\frac{1}{4}\frac{V^{\prime}}{V^{\frac{3}{2}}}\frac{1}{z^3}+1
	\right\}^2\left(\frac{1+z}{1-z}\right)^{-\frac{\sqrt{V}}{aH}}
	e^{2\frac{\sqrt{V}}{aH}z}
	+\left\{\frac{1}{4}\frac{V^{\prime}}{V^{\frac{3}{2}}}\frac{1}{z^3}-1
	\right\}^2\left(\frac{1+z}{1-z}\right)^{\frac{\sqrt{V}}{aH}}
	e^{-2\frac{\sqrt{V}}{aH}z}\right] \nonumber \\
	& =\frac{1}{4\pi^2}\frac{(V^{\prime})^2}{V}\times \frac{1}{16}
	\int_{z_0}^1dz\> \frac{\sqrt{1-z^2}}{z^4}\left\{\frac{1}{4}
	\left(\frac{1+z}{1-z}\right)^{-\frac{\sqrt{V}}{aH}}
	e^{2\frac{\sqrt{V}}{aH}z}
	+\left(\frac{1+z}{1-z}\right)^{\frac{\sqrt{V}}{aH}}
	e^{-2\frac{\sqrt{V}}{aH}z}\right\} \nonumber \\
	& \quad +\frac{1}{4\pi^2}(V^{\prime})^2\sqrt{V}\times \frac{1}{2}
	\int_{z_0}^1dz\> \frac{\sqrt{1-z^2}}{z}\left\{\frac{1}{4}
	\left(\frac{1+z}{1-z}\right)^{-\frac{\sqrt{V}}{aH}}
	e^{2\frac{\sqrt{V}}{aH}z}
	-\left(\frac{1+z}{1-z}\right)^{\frac{\sqrt{V}}{aH}}
	e^{-2\frac{\sqrt{V}}{aH}z}\right\} \nonumber \\
	& \quad +\frac{1}{4\pi^2}V^2
	\int_{z_0}^1dz\> z^2\sqrt{1-z^2}\left\{\frac{1}{4}
	\left(\frac{1+z}{1-z}\right)^{-\frac{\sqrt{V}}{aH}}
	e^{2\frac{\sqrt{V}}{aH}z}
	+\left(\frac{1+z}{1-z}\right)^{\frac{\sqrt{V}}{aH}}
	e^{-2\frac{\sqrt{V}}{aH}z}\right\} \nonumber \\
	& =\frac{a^4H^4}{4\pi^2}
	\left(4+6\frac{\dot{H}}{H^2}+\frac{\ddot{H}}{H^3}-2\frac{m^2}{H^2}
	\right)^2
	\left(2+\frac{\dot{H}}{H^2}-\frac{m^2}{H^2}\right)^{-1} \nonumber \\
	& \quad \times \Bigg[
	\frac{1}{16}\int_{z_0}^1dz\> \frac{\sqrt{1-z^2}}{z^4}
	\bigg\{\textstyle \frac{1}{4}
	\left(\frac{1+z}{1-z}\right)^{-\sqrt{2}}e^{2\sqrt{2}z}
	+\left(\frac{1+z}{1-z}\right)^{\sqrt{2}}e^{-2\sqrt{2}z}\bigg\} 
	\nonumber \\
	& \qquad \quad 
	-\frac{\sqrt{2}}{64}\left(\frac{\dot{H}}{H^2}-\frac{m^2}{H^2}\right) 
	\int_{z_0}^1dz\> \frac{\sqrt{1-z^2}}{z^4}\textstyle 
	\left(\log \frac{1+z}{1-z}-2z\right)\bigg\{\frac{1}{4}
	\left(\frac{1+z}{1-z}\right)^{-\sqrt{2}}e^{2\sqrt{2}z}
	-\left(\frac{1+z}{1-z}\right)^{\sqrt{2}}e^{-2\sqrt{2}z}\bigg\} 
	\nonumber \\
	& \qquad \quad 
	+\frac{1}{16\times 32}
	\left(\frac{\dot{H}}{H^2}-\frac{m^2}{H^2}\right)^2 \nonumber \\
	& \qquad \qquad \times 
	\int_{z_0}^1dz\> \frac{\sqrt{1-z^2}}{z^4}\textstyle 
	\bigg\{\frac{1}{4}\left((\sqrt{2}-8z)\log \frac{1+z}{1-z}
	+2\left(\log \frac{1+z}{1-z}\right)^2-2\sqrt{2}z+8z^2\right)
	\left(\frac{1+z}{1-z}\right)^{-\sqrt{2}}e^{2\sqrt{2}z} \nonumber \\
	& \qquad \qquad \qquad \qquad \qquad \qquad \textstyle 
	-\left((\sqrt{2}+8z)\log \frac{1+z}{1-z}
	-2\left(\log \frac{1+z}{1-z}\right)^2-2\sqrt{2}z-8z^2\right)
	\left(\frac{1+z}{1-z}\right)^{\sqrt{2}}e^{-2\sqrt{2}z}\bigg\} 
	\nonumber \\
	& \qquad \quad +\cdots \Bigg] \nonumber \\
	& \quad +\frac{a^4H^4}{4\pi^2}
	\left(4+6\frac{\dot{H}}{H^2}+\frac{\ddot{H}}{H^3}-2\frac{m^2}{H^2}
	\right)
	\left(2+\frac{\dot{H}}{H^2}-\frac{m^2}{H^2}\right)^{\frac{1}{2}} 
	\nonumber \\
	& \qquad \times \Bigg[
	\frac{1}{2}\int_{z_0}^1dz\> \frac{\sqrt{1-z^2}}{z}
	\bigg\{\textstyle \frac{1}{4}
	\left(\frac{1+z}{1-z}\right)^{-\sqrt{2}}e^{2\sqrt{2}z}
	-\left(\frac{1+z}{1-z}\right)^{\sqrt{2}}e^{-2\sqrt{2}z}\bigg\} 
	\nonumber \\
	& \qquad \qquad 
	-\frac{\sqrt{2}}{8}\left(\frac{\dot{H}}{H^2}-\frac{m^2}{H^2}\right) 
	\int_{z_0}^1dz\> \frac{\sqrt{1-z^2}}{z}\textstyle 
	\left(\log \frac{1+z}{1-z}-2z\right)\bigg\{\frac{1}{4}
	\left(\frac{1+z}{1-z}\right)^{-\sqrt{2}}e^{2\sqrt{2}z}
	+\left(\frac{1+z}{1-z}\right)^{\sqrt{2}}e^{-2\sqrt{2}z}\bigg\} 
	\nonumber \\
	& \qquad \qquad 
	+\frac{1}{64}
	\left(\frac{\dot{H}}{H^2}-\frac{m^2}{H^2}\right)^2 \nonumber \\
	& \qquad \qquad \quad \times 
	\int_{z_0}^1dz\> \frac{\sqrt{1-z^2}}{z}\textstyle 
	\bigg\{\frac{1}{4}\left((\sqrt{2}-8z)\log \frac{1+z}{1-z}
	+2\left(\log \frac{1+z}{1-z}\right)^2-2\sqrt{2}z+8z^2\right)
	\left(\frac{1+z}{1-z}\right)^{-\sqrt{2}}e^{2\sqrt{2}z} \nonumber \\
	& \qquad \qquad \qquad \qquad \qquad \qquad \quad \textstyle 
	+\left((\sqrt{2}+8z)\log \frac{1+z}{1-z}
	-2\left(\log \frac{1+z}{1-z}\right)^2-2\sqrt{2}z-8z^2\right)
	\left(\frac{1+z}{1-z}\right)^{\sqrt{2}}e^{-2\sqrt{2}z}\bigg\} 
	\nonumber \\
	& \qquad \qquad +\cdots \Bigg] \nonumber \\
	& \quad +\frac{a^4H^4}{4\pi^2}
	\left(2+\frac{\dot{H}}{H^2}-\frac{m^2}{H^2}\right)^2 
	\nonumber \\
	& \qquad \times \Bigg[
	\int_{z_0}^1dz\> z^2\sqrt{1-z^2}
	\bigg\{\textstyle \frac{1}{4}
	\left(\frac{1+z}{1-z}\right)^{-\sqrt{2}}e^{2\sqrt{2}z}
	+\left(\frac{1+z}{1-z}\right)^{\sqrt{2}}e^{-2\sqrt{2}z}\bigg\} 
	\nonumber \\
	& \qquad \qquad 
	-\frac{\sqrt{2}}{4}\left(\frac{\dot{H}}{H^2}-\frac{m^2}{H^2}\right) 
	\int_{z_0}^1dz\> z^2\sqrt{1-z^2}\textstyle 
	\left(\log \frac{1+z}{1-z}-2z\right)\bigg\{\frac{1}{4}
	\left(\frac{1+z}{1-z}\right)^{-\sqrt{2}}e^{2\sqrt{2}z}
	-\left(\frac{1+z}{1-z}\right)^{\sqrt{2}}e^{-2\sqrt{2}z}\bigg\} 
	\nonumber \\
	& \qquad \qquad 
	+\frac{1}{32}
	\left(\frac{\dot{H}}{H^2}-\frac{m^2}{H^2}\right)^2 \nonumber \\
	& \qquad \qquad \quad \times 
	\int_{z_0}^1dz\> z^2\sqrt{1-z^2}\textstyle 
	\bigg\{\frac{1}{4}\left((\sqrt{2}-8z)\log \frac{1+z}{1-z}
	+2\left(\log \frac{1+z}{1-z}\right)^2-2\sqrt{2}z+8z^2\right)
	\left(\frac{1+z}{1-z}\right)^{-\sqrt{2}}e^{2\sqrt{2}z} \nonumber \\
	& \qquad \qquad \qquad \qquad \qquad \qquad \quad \textstyle 
	-\left((\sqrt{2}+8z)\log \frac{1+z}{1-z}
	-2\left(\log \frac{1+z}{1-z}\right)^2-2\sqrt{2}z-8z^2\right)
	\left(\frac{1+z}{1-z}\right)^{\sqrt{2}}e^{-2\sqrt{2}z}\bigg\} 
	\nonumber \\
	& \qquad \qquad +\cdots \Bigg]. \nonumber \\
	\label{app-correlation-kinetic}
\end{align}
\begin{align}
	& \left\langle \chi^{\prime}\chi +\chi \chi^{\prime}\right\rangle 
	=\int \frac{d^3\vec{k}}{(2\pi )^3}
	\left(\chi_k^{\prime}\chi_k^{\ast}+\chi_k^{\prime \ast}\chi_k\right) 
	\nonumber \\
	& =\frac{1}{(2\pi )^3}
	\frac{2\pi^{\frac{3}{2}}}{\Gamma \left(\frac{3}{2}\right)}
	\int_0^{k_0}dk\> k^2\Bigg[
	-\frac{1}{4}\left\{\frac{V^{\prime}}{4}(V-k^2)^{-\frac{3}{2}}+1\right\}
	e^{-2\int_{\tau_k}^{\tau}d\tau^{\prime}\sqrt{V-k^2}} \nonumber \\
	& \qquad \qquad \qquad \qquad \qquad \qquad 
	-\left\{\frac{V^{\prime}}{4}(V-k^2)^{-\frac{3}{2}}-1\right\}
	e^{+2\int_{\tau_k}^{\tau}d\tau^{\prime}\sqrt{V-k^2}}\Bigg] 
	\nonumber \\
	& =-\frac{1}{2\pi^2}\int_0^{k_0}dk\> k^2
	\Bigg[\frac{1}{4}\left\{\frac{V^{\prime}}{4}(V-k^2)^{-\frac{3}{2}}+1
	\right\}
	\left(\textstyle \frac{\sqrt{V}+\sqrt{V-k^2}}{\sqrt{V}-\sqrt{V-k^2}}
	\right)^{-\frac{\sqrt{V}}{aH}}e^{2\frac{\sqrt{V-k^2}}{aH}} \nonumber \\
	& \qquad \qquad \qquad \qquad 
	+\left\{\frac{V^{\prime}}{4}(V-k^2)^{-\frac{3}{2}}-1\right\}
	\left(\textstyle \frac{\sqrt{V}+\sqrt{V-k^2}}{\sqrt{V}-\sqrt{V-k^2}}
	\right)^{\frac{\sqrt{V}}{aH}}e^{-2\frac{\sqrt{V-k^2}}{aH}}\Bigg] 
	\nonumber \\
	& =-\frac{V^{\frac{3}{2}}}{2\pi^2}
	\int_{z_0}^1dz\> z\sqrt{1-z^2}\left[\frac{1}{4}
	\left(\frac{1}{4}\frac{V^{\prime}}{V^{\frac{3}{2}}}\frac{1}{z^3}+1
	\right)\left(\frac{1+z}{1-z}\right)^{-\frac{\sqrt{V}}{aH}}
	e^{2\frac{\sqrt{V}}{aH}z}
	+\left(\frac{1}{4}\frac{V^{\prime}}{V^{\frac{3}{2}}}\frac{1}{z^3}-1
	\right)\left(\frac{1+z}{1-z}\right)^{\frac{\sqrt{V}}{aH}}
	e^{-2\frac{\sqrt{V}}{aH}z}\right] \nonumber \\
	& =-\frac{1}{4\pi^2}V^{\prime}\times \frac{1}{2}
	\int_{z_0}^1dz\> \frac{\sqrt{1-z^2}}{z^2}\left\{\frac{1}{4}
	\left(\frac{1+z}{1-z}\right)^{-\frac{\sqrt{V}}{aH}}
	e^{2\frac{\sqrt{V}}{aH}z}
	+\left(\frac{1+z}{1-z}\right)^{\frac{\sqrt{V}}{aH}}
	e^{-2\frac{\sqrt{V}}{aH}z}\right\} \nonumber \\
	& \quad -\frac{1}{4\pi^2}V^{\frac{3}{2}}\times 2
	\int_{z_0}^1dz\> z\sqrt{1-z^2}\left\{\frac{1}{4}
	\left(\frac{1+z}{1-z}\right)^{-\frac{\sqrt{V}}{aH}}
	e^{2\frac{\sqrt{V}}{aH}z}
	-\left(\frac{1+z}{1-z}\right)^{\frac{\sqrt{V}}{aH}}
	e^{-2\frac{\sqrt{V}}{aH}z}\right\} \nonumber \\
	& =-\frac{a^3H^3}{4\pi^2}
	\left(4+6\frac{\dot{H}}{H^2}+\frac{\ddot{H}}{H^3}-2\frac{m^2}{H^2}
	\right) \nonumber \\
	& \quad \times \Bigg[
	\frac{1}{2}\int_{z_0}^1dz\> \frac{\sqrt{1-z^2}}{z^2}
	\bigg\{\textstyle \frac{1}{4}
	\left(\frac{1+z}{1-z}\right)^{-\sqrt{2}}e^{2\sqrt{2}z}
	+\left(\frac{1+z}{1-z}\right)^{\sqrt{2}}e^{-2\sqrt{2}z}\bigg\} 
	\nonumber \\
	& \qquad \quad 
	-\frac{\sqrt{2}}{8}\left(\frac{\dot{H}}{H^2}-\frac{m^2}{H^2}\right) 
	\int_{z_0}^1dz\> \frac{\sqrt{1-z^2}}{z^2}\textstyle 
	\left(\log \frac{1+z}{1-z}-2z\right)\bigg\{\frac{1}{4}
	\left(\frac{1+z}{1-z}\right)^{-\sqrt{2}}e^{2\sqrt{2}z}
	-\left(\frac{1+z}{1-z}\right)^{\sqrt{2}}e^{-2\sqrt{2}z}\bigg\} 
	\nonumber \\
	& \qquad \quad 
	+\frac{1}{64}
	\left(\frac{\dot{H}}{H^2}-\frac{m^2}{H^2}\right)^2 \nonumber \\
	& \qquad \qquad \times 
	\int_{z_0}^1dz\> \frac{\sqrt{1-z^2}}{z^2}\textstyle 
	\bigg\{\frac{1}{4}\left((\sqrt{2}-8z)\log \frac{1+z}{1-z}
	+2\left(\log \frac{1+z}{1-z}\right)^2-2\sqrt{2}z+8z^2\right)
	\left(\frac{1+z}{1-z}\right)^{-\sqrt{2}}e^{2\sqrt{2}z} \nonumber \\
	& \qquad \qquad \qquad \qquad \qquad \qquad \textstyle 
	-\left((\sqrt{2}+8z)\log \frac{1+z}{1-z}
	-2\left(\log \frac{1+z}{1-z}\right)^2-2\sqrt{2}z-8z^2\right)
	\left(\frac{1+z}{1-z}\right)^{\sqrt{2}}e^{-2\sqrt{2}z}\bigg\} 
	\nonumber \\
	& \qquad \quad +\cdots \Bigg] \nonumber \\
	& \quad -\frac{a^3H^3}{4\pi^2}
	\left(2+\frac{\dot{H}}{H^2}-\frac{m^2}{H^2}\right)^{\frac{3}{2}} 
	\nonumber \\
	& \qquad \times \Bigg[
	2\int_{z_0}^1dz\> z\sqrt{1-z^2}
	\bigg\{\textstyle \frac{1}{4}
	\left(\frac{1+z}{1-z}\right)^{-\sqrt{2}}e^{2\sqrt{2}z}
	-\left(\frac{1+z}{1-z}\right)^{\sqrt{2}}e^{-2\sqrt{2}z}\bigg\} 
	\nonumber \\
	& \qquad \qquad 
	-\frac{\sqrt{2}}{2}\left(\frac{\dot{H}}{H^2}-\frac{m^2}{H^2}\right) 
	\int_{z_0}^1dz\> z\sqrt{1-z^2}\textstyle 
	\left(\log \frac{1+z}{1-z}-2z\right)\bigg\{\frac{1}{4}
	\left(\frac{1+z}{1-z}\right)^{-\sqrt{2}}e^{2\sqrt{2}z}
	+\left(\frac{1+z}{1-z}\right)^{\sqrt{2}}e^{-2\sqrt{2}z}\bigg\} 
	\nonumber \\
	& \qquad \qquad 
	+\frac{1}{16}
	\left(\frac{\dot{H}}{H^2}-\frac{m^2}{H^2}\right)^2 \nonumber \\
	& \qquad \qquad \quad \times 
	\int_{z_0}^1dz\> z\sqrt{1-z^2}\textstyle 
	\bigg\{\frac{1}{4}\left((\sqrt{2}-8z)\log \frac{1+z}{1-z}
	+2\left(\log \frac{1+z}{1-z}\right)^2-2\sqrt{2}z+8z^2\right)
	\left(\frac{1+z}{1-z}\right)^{-\sqrt{2}}e^{2\sqrt{2}z} \nonumber \\
	& \qquad \qquad \qquad \qquad \qquad \qquad \quad \textstyle 
	+\left((\sqrt{2}+8z)\log \frac{1+z}{1-z}
	-2\left(\log \frac{1+z}{1-z}\right)^2-2\sqrt{2}z-8z^2\right)
	\left(\frac{1+z}{1-z}\right)^{\sqrt{2}}e^{-2\sqrt{2}z}\bigg\} 
	\nonumber \\
	& \qquad \qquad +\cdots \Bigg].
	\label{app-correlation-cross}
\end{align}



\end{document}